\documentclass[twoside,australian,sort&compress]{iopart}
\usepackage[T1]{fontenc}
\usepackage[utf8]{inputenc}
\usepackage{geometry}
\usepackage{float}
\usepackage{graphicx}
\usepackage[numbers]{natbib}

\makeatletter

\providecommand{\tabularnewline}{\\}

\usepackage{iopams}
\usepackage{setstack}

\newcommand{\eqref}[1]{(\ref{#1})}

\usepackage{xurl}

\makeatother

\usepackage{babel}
\begin{document}
\title{An improved set of electron-THFA cross sections refined through a
neural network-based analysis of swarm data}
\author{{\Large{}P W Stokes$^{1}$, S P Foster$^{1}$, M J E Casey$^{1}$,
D G Cocks$^{2}$, O González-Magaña$^{3}$, J de Urquijo$^{3}$, G
García$^{4}$, M J Brunger$^{5,6}$ and R D White$^{1}$}}
\address{{\Large{}$^{1}$}{\large{}College of Science and Engineering, James
Cook University, Townsville, QLD 4811, Australia}}
\address{{\Large{}$^{2}$}{\large{}Research School of Physics, Australian National
University, Canberra, ACT 0200, Australia}}
\address{{\Large{}$^{3}$}{\large{}Instituto de Ciencias Físicas, Universidad
Nacional Autónoma de México, 62251, Cuernavaca, Morelos, Mexico}}
\address{{\Large{}$^{4}$}{\large{}Instituto de Física Fundamental, CSIC, Serrano
113-bis, 28006 Madrid, Spain}}
\address{{\Large{}$^{5}$}{\large{}College of Science and Engineering, Flinders
University, Bedford Park, Adelaide, SA 5042, Australia}}
\address{{\Large{}$^{6}$}{\large{}Department of Actuarial Science and Applied
Statistics, Faculty of Business and Management, UCSI University, Kuala
Lumpur 56000, Malaysia}}
\ead{{\large{}peter.stokes@my.jcu.edu.au}}
\begin{abstract}
We review experimental and theoretical cross sections for electron
transport in $\alpha$-tetrahydrofurfuryl alcohol (THFA) and, in doing
so, propose a plausible complete set. To assess the accuracy and self-consistency
of our proposed set, we use the pulsed-Townsend technique to measure
drift velocities, longitudinal diffusion coefficients and effective
Townsend first ionisation coefficients for electron swarms in admixtures
of THFA in argon, across a range of density-reduced electric fields
from 1 Td to 450 Td. These measurements are then compared to simulated
values derived from our proposed set using a multi-term solution of
Boltzmann's equation. We observe discrepancies between the simulation
and experiment, which we attempt to address by employing a neural
network model that is trained to solve the inverse swarm problem of
unfolding the cross sections underpinning our experimental swarm measurements.
What results from our neural network-based analysis is a refined set
of electron-THFA cross sections, which we confirm is of higher consistency
with our swarm measurements than that we initially proposed. We also
use our data base to calculate electron transport coefficients in
pure THFA, across a range of reduced electric fields from 0.001 Td
to 10,000 Td. 
\end{abstract}
\noindent{\it Keywords\/}: {swarm analysis, inverse problem, Boltzmann equation, machine learning,
$\alpha$-tetrahydrofurfuryl alcohol\\
}
\submitto{\JCP }
\maketitle

\section{\label{sec:Introduction}Introduction}

The study of non-equilibrium electron transport in biological matter
underpins a diverse range of scientific fields and applications. Of
particular interest is the medical sector, where electron-induced
processes in human tissue occur in both medical imaging and therapy
\citep{gomez2012radiation}. In these applications, ionising radiation
liberates large numbers of low-energy secondary electrons ($\sim30\ \mathrm{eV}$)
which undergo a variety of energy deposition processes in the biomolecules
that constitute human tissue \citep{Boudaiffa2000}. These thermalised
electrons are known to undergo dissociative electron attachment (DEA),
which has been attributed in part to the damage associated with such
ionising radiation, either directly through inducing single or double
strand breaks in DNA, or indirectly through the interactions of electron-induced
radicals with DNA. Accurate kinetic simulations of electron transport
in biological matter, including a full description of the interactions
with each of the various biomolecular constituents, are therefore
required in order to fully understand radiation damage and comprehensively
inform dosimetry models.

The modelling of electron transport in biomolecules has also found
recent application in plasma medicine, which is a relatively new field
motivated by the synergistic interactions of low-temperature atmospheric
pressure plasmas (LTAPP) with biological tissue \citep{Kong2009,Bruggeman2016,Adamovich2017}.
Human tissue is generally modelled as a bulk liquid so that the system
simplifies to a three-phase problem consisting of a bulk gas, a gas-liquid
interface, and a bulk liquid \citep{Bruggeman2016,Adamovich2017}.
While the interactions of the reactive oxygen and nitrogen species
(RONS), produced in the plasma-liquid interface, are known to induce
many of the synergistic effects \citep{Kong2009}, a predictive understanding
of plasma treatments can only be obtained through a complete understanding
of all of the plasma-tissue interactions, including the electron-impact
generation of radicals \citep{Boudaiffa2000}.

Despite its importance, a full soft-condensed phase tissue description
of electron transport is currently in its infancy. When applying kinetic
modelling techniques, biological matter is currently approximated
as water vapour, despite biological media being neither water nor
a gas, nor being decoupled entirely from electron interactions as
in plasma medical device modelling, but rather a complex mixture of
biological molecules in a soft-condensed phase \citep{White2014,Nikjoo2006,Francis2011}.
As such, quantitative modelling of electron transport through biological
media requires the attainment of complete and accurate sets of cross
sections for all electron interactions with all relevant biomolecules,
including water, in the soft-condensed phase.

As low-energy electron interactions with DNA are difficult to study,
focus has turned to the individual components of DNA, in addition
to their structural analogues. One component that has received considerable
attention is 2-deoxyribose, a sugar that links phosphate groups in
the DNA backbone, which has well-studied surrogates including tetrahydrofuran
(THF, $\mathrm{C}_{4}\mathrm{H}_{8}\mathrm{O}$) and $\alpha$-tetrahydrofurfuryl
alcohol (THFA, $\mathrm{C_{5}H_{10}O_{2}}$) \citep{Brunger2017}.
Between these, THF has received the most attention, with a number
of proposed complete electron impact cross section sets present in
the literature \citep{Garland2013,Fuss2014,Bug2017,Swadia2017,Swadia2017a,Casey2017,DeUrquijo2019a,Stokes2020}.
In comparison, however, while individual electron-THFA cross sections
are known, a complete set is presently still lacking. In this investigation,
we attempt to remedy this gap in the literature by constructing and
refining a complete and self-consistent set of electron-THFA cross
sections in the gas-phase, with the motivation being that such a set
can be adapted to the soft-condensed phase through appropriate modifications
using pair correlations functions \citep{White2009a,White2014a}.
The present investigation is especially warranted given that, in comparison
to THF, THFA has been identified as a superior analogue for 2-deoxyribose
\citep{Limao-Vieira2014,Duque2014a}.

Key to the derivation of our cross section set is the measurement
and subsequent analysis of electron swarm transport coefficients in
admixtures of THFA and argon. By comparing these measurements to simulated
transport coefficients, the accuracy and self-consistency of our cross
section set can be assessed \citep{White2018}. Any discrepancies
that are observed can then be used to inform appropriate adjustments
to the cross sections in order to reduce the discrepancies with experiment.
By iterating this process, such discrepancies can be minimised. This
\textit{inverse swarm problem} of unfolding cross sections from swarm
measurements has a long and successful history \citep{Mayer1921,Ramsauer1921,Townsend1922,Frost1962,Engelhardt1963,Engelhardt1964,Hake1967,Phelps1968}.
However, being an inverse problem, it is often the case that there
is no single unique set of cross sections that is consistent with
a given set of swarm measurements. This nonuniqueness poses a fundamental
challenge in automating swarm analysis using numerical optimisation
algorithms \citep{Duncan1972,OMalley1980,Taniguchi1987,Suzuki1989,Suzuki1990,Morgan1991a,Morgan1993,Brennan1993},
as while such algorithms diligently minimise the error in the associated
transport coefficients, they lack the intuition about what constitutes
a physically-plausible cross section set. As such, to try and ensure
the most success, these iterative adjustments to the cross section
set must be carefully performed by an expert that can use their prior
knowledge to rule out unphysical solutions. Despite these challenges,
we have recently had some success in employing machine learning models
to solve the inverse swarm problem automatically \citep{Stokes2019,Stokes2020},
an approach that was originally explored by Morgan \citep{Morgan1991}
decades earlier. By training these models on cross sections derived
from the LXCat project \citep{Pancheshnyi2012,Pitchford2017,LXCat},
they can, in a sense, ``learn'' what constitutes a physically-plausible
cross section set. Recently \citep{Stokes2020}, we trained an artificial
neural network model in order to refine the electron-THF cross section
set of de Urquijo \textit{et al}. \citep{DeUrquijo2019a}. Promisingly,
the set of cross sections determined by this neural network was found
to be of comparable quality to the de Urquijo \textit{et al}. set
that was refined ``by hand''. LXCat cross sections have also been
applied recently by Nam \textit{et al}. \citep{Nam2021} to train
a neural network for the classification of cross sections according
to their type (i.e. elastic, excitation, ionisation, or attachment).

The remainder of this paper is structured as follows. In Section \ref{sec:Neural-network-for},
we briefly describe our data-driven approach to solving the inverse
swarm problem, including the nature of the cross sections and transport
coefficients used to train the machine learning model. Section \ref{sec:Electron-THFA-cross-section}
provides a review of existing electron-THFA cross sections in the
literature. These measured and calculated integral cross sections
(ICSs) are then employed to construct a ``proposed'' data base for
electron-THFA scattering, which is also described in this section.
Note that several of the present authors have had recent experience
in constructing data bases for electron and positron scattering problems
\citep{McEachran2020,Blanco2019,McEachran2018,Brunger2002}, but that
the success of this approach does depend on the volume of relevant
data available and by its very construction can be highly selective.
In Section \ref{sec:Transport-coefficients-of-PROPOSED} details of
our experimental technique for measuring the THFA-argon gas mixture
transport coefficients are provided, with the results of these measurements
also being presented. Note also in this section that results from
our Boltzmann equation analysis, using our ``proposed'' cross sections,
are provided and compared against the measured data. In Section \ref{sec:Refined-set-of}
a refined set of THFA electron scattering cross sections, using our
machine learning / neural network-based approach, are presented and
discussed, with results from their application in our Boltzmann equation
analysis, for simulated transport parameters, being given in Section
\ref{sec:Transport-coefficients-of-REFINED}. Finally, Section \ref{sec:Conclusion}
presents our conclusions from the current investigation and gives
some suggestions for future work. 

\section{\label{sec:Neural-network-for}Neural network for cross section regression}

In this section, we briefly describe the architecture and application
of our neural network for the regression of electron-THFA cross sections
from swarm transport data. For a more detailed description of this
approach to inverting the swarm problem, we refer the reader to our
previous publications, Refs. \citep{Stokes2019} and \citep{Stokes2020}.

\subsection{Neural network architecture}

In this work, we perform the cross section regression by utilising
neural networks of the form:
\begin{equation}
\mathbf{y}\left(\mathbf{x}\right)=\left(\mathbf{A}_{4}\circ\mathrm{mish}\circ\mathbf{A}_{3}\circ\mathrm{mish}\circ\mathbf{A}_{2}\circ\mathrm{mish}\circ\mathbf{A}_{1}\right)\left(\mathbf{x}\right),\label{eq:nn}
\end{equation}
where $\mathbf{A}_{n}\left(\mathbf{x}\right)\equiv\mathbf{W}_{n}\mathbf{x}+\mathbf{b}_{n}$
are affine mappings, $\circ$ denotes function composition, and $\mathrm{mish}\left(x\right)=x\tanh\left(\ln\left(1+e^{x}\right)\right)$
\citep{Misra2019} is a nonlinear \textit{activation function} that
is applied element-wise. The neural network, Eq. \eqref{eq:nn}, is
said to be \textit{fully-connected} as the matrices of \textit{weights},
$\mathbf{W}_{n}$, and vectors of \textit{biases}, $\mathbf{b}_{n}$,
are dense. The number of weights and biases is correlated with the
\textit{capacity} of the neural network to perform a particular nonlinear
mapping from the input vector $\mathbf{x}$ to the output vector $\mathbf{y}$.
With the exception of $\mathbf{b}_{4}$, the size of which must match
the output of the network, we specify 256 biases per bias vector and
size the weight matrices accordingly. Naturally, the output of our
neural network for swarm analysis contains the electron-THFA cross
sections of interest:
\begin{equation}
\mathbf{y}=\left[\begin{array}{c}
\sigma_{1}\left(\varepsilon\right)\\
\sigma_{2}\left(\varepsilon\right)\\
\vdots
\end{array}\right].
\end{equation}
As these cross sections are functions of energy, we accordingly include
the energy $\varepsilon$ as an element of the input vector $\mathbf{x}$.
To solve the inverse swarm problem, we populate the remaining input
elements with the swarm transport coefficient measurements:
\begin{equation}
\mathbf{x}=\left[\begin{array}{c}
\varepsilon\\
\hline W_{1}\\
W_{2}\\
\vdots\\
\hline \left(\alpha_{\mathrm{eff}}/n_{0}\right)_{1}\\
\left(\alpha_{\mathrm{eff}}/n_{0}\right)_{2}\\
\vdots\\
\hline \left(n_{0}D_{L}\right)_{1}\\
\left(n_{0}D_{L}\right)_{2}\\
\vdots
\end{array}\right],
\end{equation}
where $W$ denotes the drift velocity, $\alpha_{\mathrm{eff}}$ denotes
the effective Townsend first ionisation coefficient, $D_{L}$ denotes
the longitudinal diffusion coefficient, and $n_{0}$ is the background
neutral number density.

Cross section regression in this way is particularly appealing due
to the versatility of neural networks. So long as the energy $\varepsilon$
remains an input to the network, we can in principle derive cross
sections from any collection of experimental measurements. In fact,
in forming our initial proposed set of electron-THFA cross sections
in Section \ref{sec:Electron-THFA-cross-section}, we derive plausible
vibrational and electronic excitation cross sections in this way from
the limited number of experimental measurements that are currently
available.

It should lastly be noted that we normalise all inputs and outputs
of the neural network by first taking the logarithm and then performing
a linear mapping onto the domain $\left[-1,1\right]$. As a consequence
of this log-transformation, we are restricted to inputs and outputs
that are positive. This poses a difficulty when predicting cross sections
below threshold. In such instances, we replace cross sections equal
to zero with a suitably small positive number, which we take to be
$10^{-26}\ \mathrm{m}^{2}$. Consequently, if the neural network outputs
a cross section less than $10^{-26}\ \mathrm{m}^{2}$, we interpret
the output as zero. This allows the threshold energy to be determined
implicitly from the output of the neural network.

\subsection{Training data}

For training the neural network, Eq. \eqref{eq:nn}, we use, as required,
elastic, excitation, ionisation and attachment cross sections from
the LXCat project \citep{Pancheshnyi2012,Pitchford2017,LXCat,Biagi,Biagiv71,Bordage,BSR,CCC,Christophorou,COP,eMolLeHavre,FLINDERS,Hayashi,ISTLisbon,Itikawa,Morgan,NGFSRDW,Phelps,Puech,QUANTEMOL,SIGLO,TRINITI}.
Specifically, we generate realistic cross sections for training by
taking random pairwise geometric combinations of cross sections from
LXCat using the formula:
\begin{equation}
\sigma\left(\varepsilon\right)=\sigma_{1}^{1-r}\left(\varepsilon+\varepsilon_{1}-\varepsilon_{1}^{1-r}\varepsilon_{2}^{r}\right)\sigma_{2}^{r}\left(\varepsilon+\varepsilon_{2}-\varepsilon_{1}^{1-r}\varepsilon_{2}^{r}\right),\label{eq:mixture}
\end{equation}
where $\sigma_{1}\left(\varepsilon\right)$ and $\sigma_{2}\left(\varepsilon\right)$
are cross sections of a given process chosen randomly without replacement,
$\varepsilon_{1}$ and $\varepsilon_{2}$ are their respective threshold
energies, and $r\in\left[0,1\right]$ is a uniformly sampled mixing
ratio. We apply similar geometric combinations when we wish to constrain
training cross sections within known experimental error bars. Specifically,
we ensure $\sigma_{1}\left(\varepsilon\right)$ is itself constrained
and then perturb about it with a random $\sigma_{2}\left(\varepsilon\right)$
and a mixing ratio $r$ chosen small enough so as not to violate the
prescribed constraints.

Once cross sections have been selected for training, they must be
sampled at various energies within the domain of interest. In this
work, we are concerned with the domain $\varepsilon\in\left[10^{-3}\ \mathrm{eV},10^{3}\ \mathrm{eV}\right]$,
which we sample randomly within using:
\begin{equation}
\varepsilon=10^{s}\ \mathrm{eV},\label{eq:sample}
\end{equation}
where $s\in\left[-3,3\right]$ is a uniformly distributed random number.

To complete the input vector of our input/output training pair, we
calculate corresponding transport coefficients using a well-benchmarked
multi-term solution of Boltzmann's equation \citep{White2009,Boyle2017,White2018}.
For good measure, we employ the ten-term approximation for all cross
section sets used for training. Additionally, to simulate the random
error present in the experimental swarm measurements, we multiply
our simulated transport coefficients by a small amount of random noise
sampled from a log-normal distribution. Specifically, we sample the
natural logarithm of this noise factor from a normal distribution
with a mean of $0$ and a standard deviation of $0.03$.

\subsection{Training procedure}

The neural network is implemented and trained using the \textit{Flux.jl}
machine learning framework \citep{Innes2018}. Before training, we
initialise the neural network biases to zero and weights to uniform
random numbers as described by Glorot and Bengio \citep{Glorot2010}.
Then, to train the network, we perform numerical optimisation of its
weights and biases so as to minimise the mean absolute error of the
cross sections fitted by the neural network. We choose the mean absolute
error measure due to its robustness in the presence of outliers. During
the optimisation we repeatedly update the weights and biases using
the Adam optimiser \citep{Kingma2015} with step size $\alpha=10^{-3}$,
exponential decay rates $\beta_{1}=0.9$ and $\beta_{2}=0.999$, and
small parameter $\epsilon=10^{-8}$. For each update of the neural
network parameters with the optimiser, we consider a random batch
of $4096$ input/output training examples. Each of these batches consist
of $16$ random LXCat-derived cross section sets, with each set sampled
at $256$ random energies using Eq. \eqref{eq:sample}. In total,
the training data set consists of 50,000 unique sets of cross sections.
Training is continued until the transport coefficients resulting from
the fitted cross section set best match the measured pulsed-Townsend
transport coefficients that were used to perform the fit.

\section{\label{sec:Electron-THFA-cross-section}Electron-THFA cross section
data review and initial proposals}

\subsection{\label{subsec:Electronic-excitation-cross}Electronic excitation
cross sections}

\begin{figure}
\begin{centering}
\includegraphics[scale=0.5]{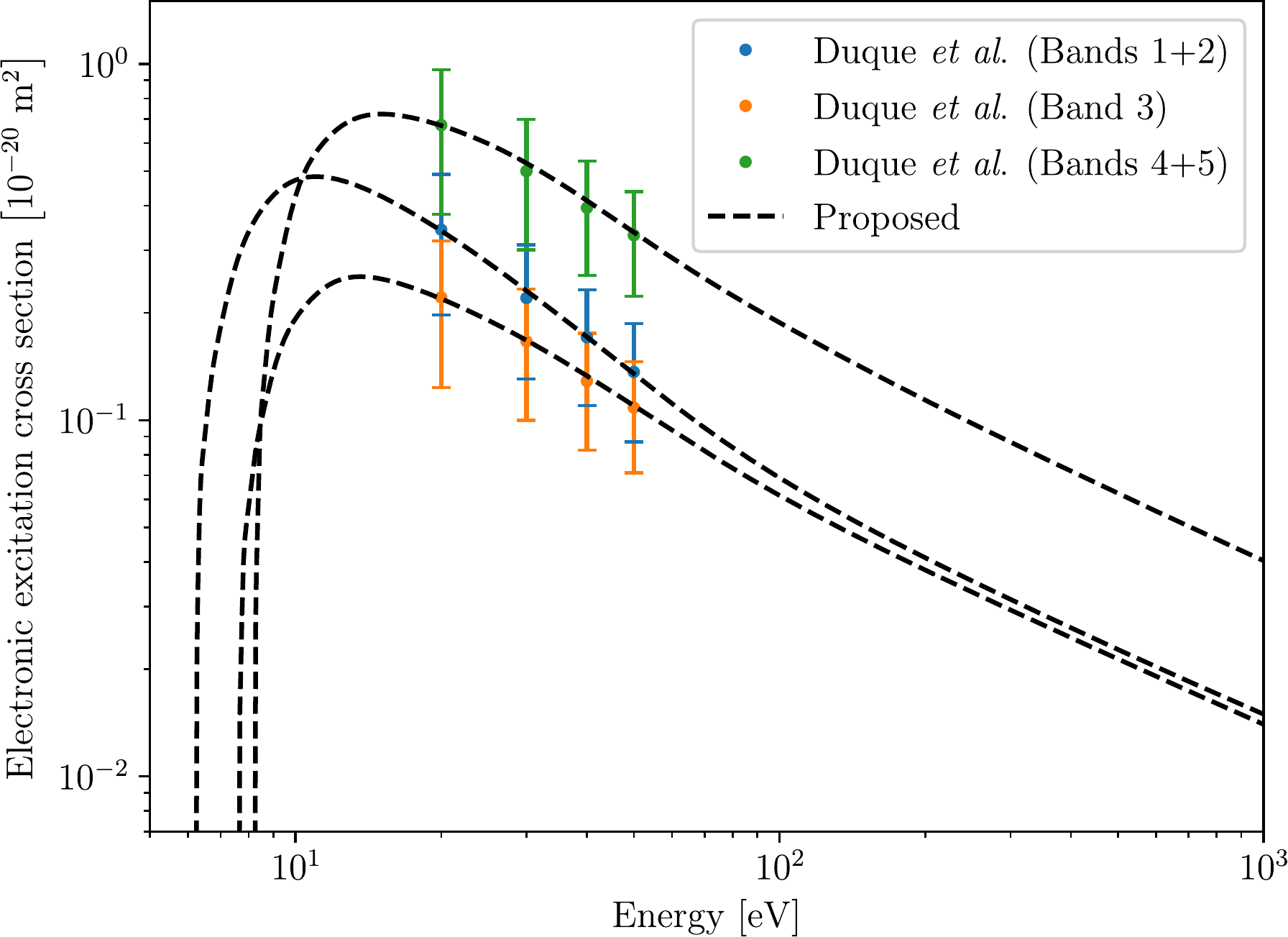}
\par\end{centering}
\caption{\label{fig:Electronic-cross-section}Proposed THFA discrete electronic
excitation cross sections, alongside the experimental measurements
\citep{Duque2014a} to which our initial proposals were fitted. See
also legend in figure.}
\end{figure}

Electronic-state scattering results for electron-THFA includes the
experimental differential cross section (DCS) data of Chiari \textit{et
al}. \citep{Chiari2014} and the corresponding integral cross section
(ICS) data of Duque \textit{et al}. \citep{Duque2014a}, the latter
of which are plotted in Figure \ref{fig:Electronic-cross-section}.
These authors identify five Rydberg electronic-state bands for THFA
\citep{Limao-Vieira2014}, although due to insufficient energy resolution
in the experimental apparatus these are resolved into only three separate
electronic-state bands. These bands are reported as having threshold
energies of 6.2 eV, 7.6 eV and 8.2 eV for Bands 1+2, Band 3 and Bands
4+5, respectively. To construct our proposed electronic excitation
cross sections for THFA, we make use of these threshold energies alongside
the ICS data of Duque \textit{et al}. To interpolate this data, as
well as extrapolate to higher energies, we employ a neural network
of the form of Eq. \eqref{eq:nn} to fit a plausible excitation cross
section for each case. Specifically, for a given band, we input to
the network the ICS data at the four energies considered by Chiari
\textit{et al}. and Duque \textit{et al}. (20 eV, 30 eV, 40 eV and
50 eV), as well as the threshold energy for that band. The resulting
neural network regression, and initial proposed ICS, for each band
is plotted in Figure \ref{fig:Electronic-cross-section}. In each
case, a plausible energy location of the peak cross section value
is identified automatically by the network.

\subsection{Vibrational excitation cross sections}

\begin{figure}
\begin{centering}
\includegraphics[scale=0.5]{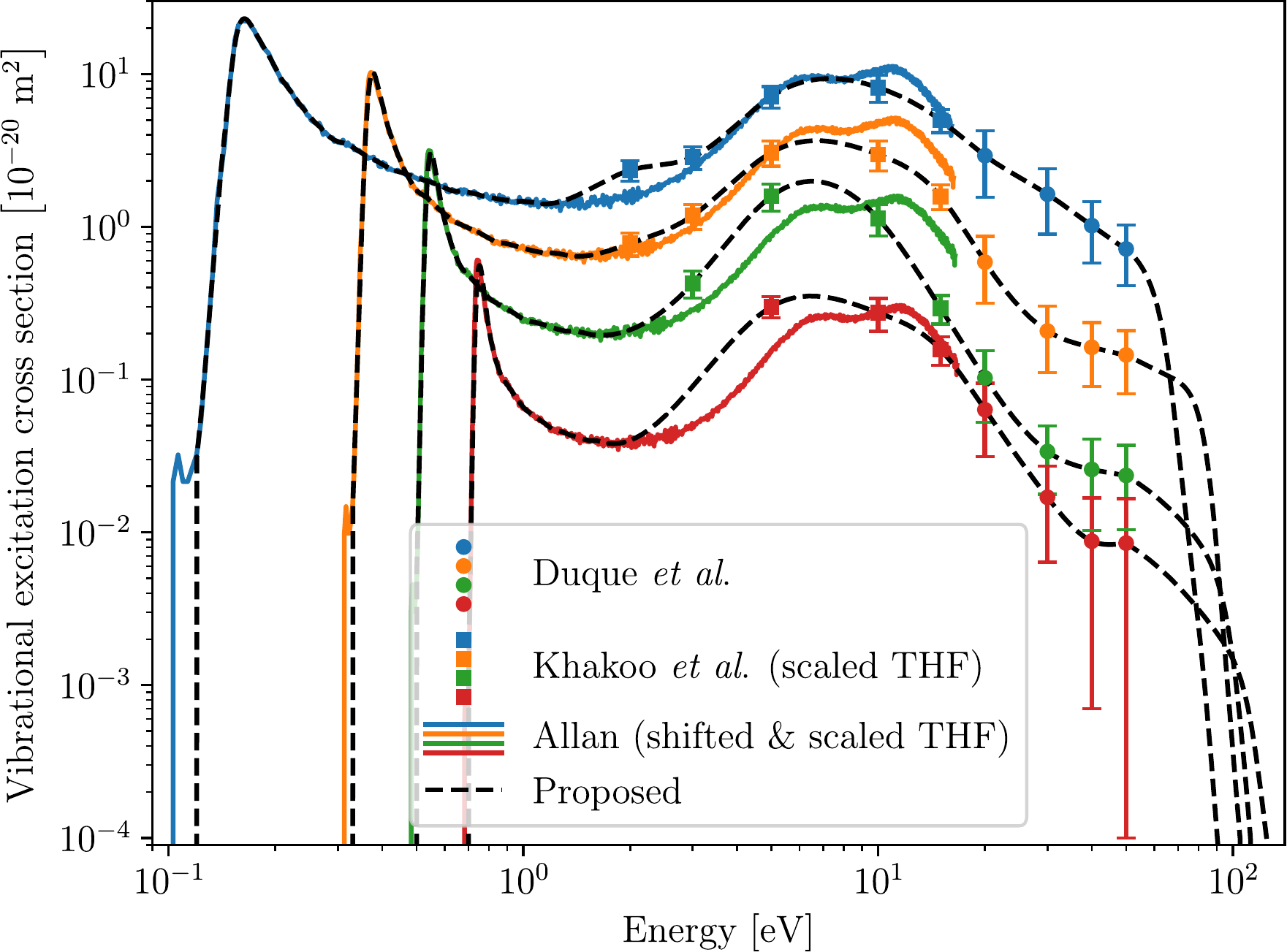}
\par\end{centering}
\caption{\label{fig:Vibrational-cross-section}Proposed vibrational excitation
cross sections, alongside the experimental measurements \citep{Duque2014,Khakoo2013,Allan2007}
to which our initial proposed data were fitted. The four vibrational
modes here are $\mathrm{CC}$ stretch (blue), $\mathrm{CH_{2}}$ stretch
(orange), $\mathrm{OH}$ stretch + combination (green), and $2\times\mathrm{CH_{2}}$
stretch (red). See also legend in figure.}
\end{figure}

Vibrational scattering data for electron-THFA consists only of the
experimental DCS and ICS data of Duque \textit{et al}. \citep{Duque2014},
at the incident energies of 20 eV, 30 eV, 40 eV and 50 eV. The four
vibrational modes of THFA identified by Duque \textit{et al}. are
the $\mathrm{CC}$ stretch, $\mathrm{CH_{2}}$ stretch, $\mathrm{OH}$
stretch + combination band, and $2\times\mathrm{CH_{2}}$ stretch
overtone, with respective threshold energies of approximately 0.12
eV, 0.33 eV, 0.5 eV and 0.7 eV. For our proposed THFA vibrational
excitation cross sections, we make use of this ICS data with these
corresponding energy thresholds. At lower energies, we employ the
THF data of Khakoo \textit{et al}. \citep{Khakoo2013}, for the same
vibrational modes, defined at the energies 2 eV, 3 eV, 5 eV, 10 eV,
15 eV and 20 eV. Note that here we scale the Khakoo \textit{et al}.
data a little, so as to match the THFA data of Duque \textit{et al}.
at the overlapping point of 20 eV. This approach is thought to be
reasonable due to THF and THFA having similar structures and intrinsic
molecular properties (e.g. dipole moment and dipole polarisability).
At very low energies, down to the threshold in each case, we make
use of the THF CC stretch data of Allan \citep{Allan2007}. To accomplish
this we need to shift Allan's measurement to each respective THFA
threshold, and also to scale so as to minimise the discontinuity with
the overlapping scaled measurements of Khakoo \textit{et al}. We then
subsequently performed a smoothing cubic spline interpolation \citep{Reinsch1967}
through the measurements of Duque \textit{et al}., the scaled measurements
of Khakoo \textit{et al}., and the shifted and scaled measurement
of Allan \textit{et al}. (up to 1 eV above threshold in each case).
Finally, above 50 eV we employ the same neural network regression
approach used in the previous section for electronic excitation interpolation/extrapolation.
Specifically, for each vibrational mode, we consider as input to that
neural network the threshold energy and the four measurements of Duque
\textit{et al}. in each case. The output of the neural network is
then the vibrational integral cross section at points above 50 eV
(up to 1000 eV). We join the resulting high-energy extrapolation with
the smoothing cubic spline interpolation at 50 eV by a further scaling.
The final proposed vibrational excitation cross sections for THFA
are plotted in Figure \ref{fig:Vibrational-cross-section}, alongside
all of the experimental measurements from which they are derived.

\subsection{Electron impact ionisation cross section}

\begin{figure}
\begin{centering}
\includegraphics[scale=0.5]{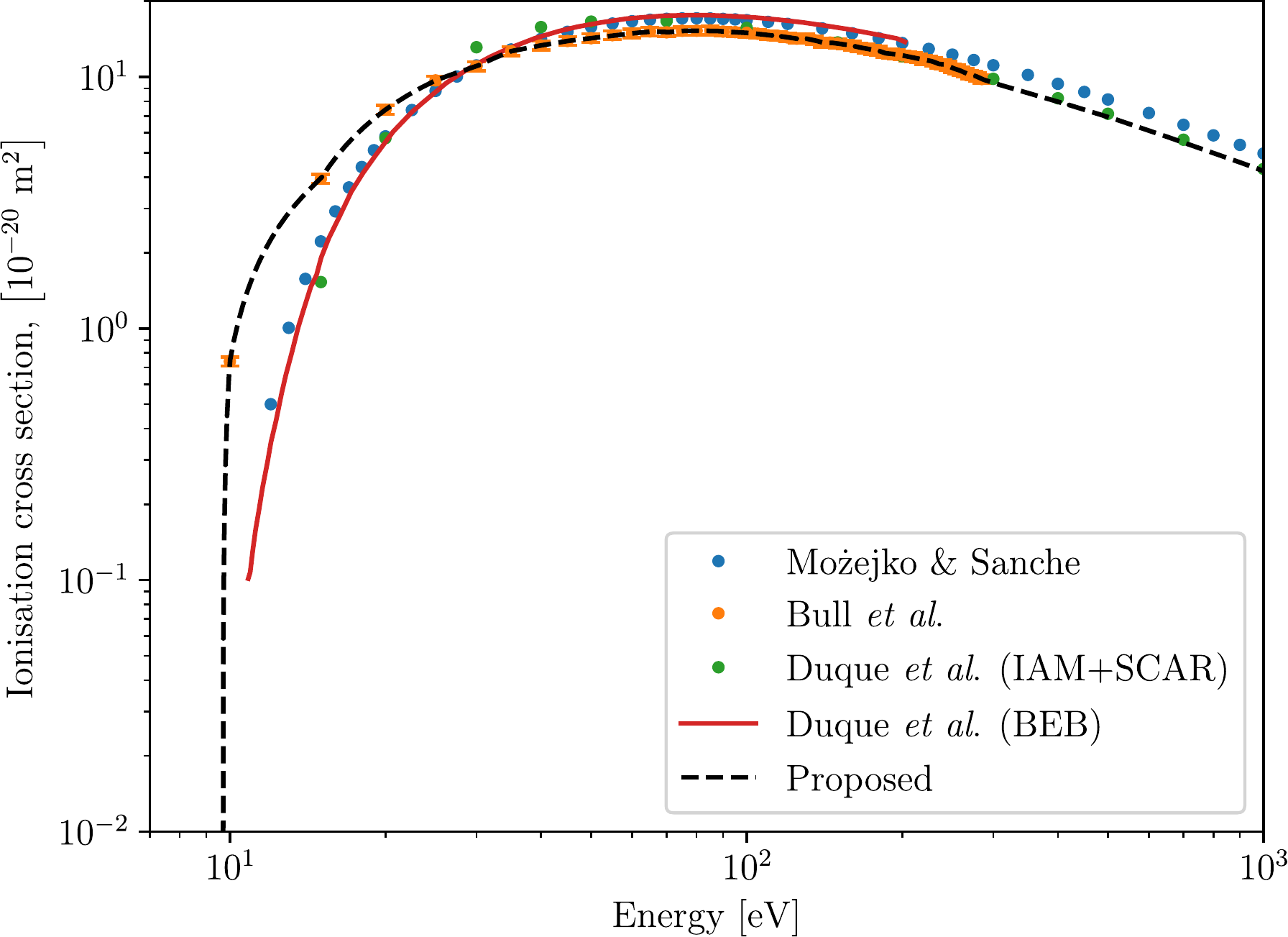}
\par\end{centering}
\caption{\label{fig:Ionisation-cross-section}Proposed electron impact ionisation
cross section, alongside previous experimental and theoretical results
\citep{Mozejko2005,Bull2014,Duque2014a} from which it was derived.
See also legend in figure.}
\end{figure}
Available scattering data for electron impact ionisation of THFA includes
the theoretical ICS data of Możejko and Sanche \citep{Mozejko2005},
the experimental ICS results of Bull \textit{et al}. \citep{Bull2014},
and the theoretical ICS data of Duque \textit{et al}. \citep{Duque2014a}.
These are each plotted in Figure \ref{fig:Ionisation-cross-section}.
Both of the aforementioned theoretical investigations use a semi-classical
binary-encounter-Bethe (BEB) formalism \citep{Mozejko2005,Duque2014a,Tanaka2016},
with Duque \textit{et al}. also employing a modified IAM-SCAR approach
\citep{Duque2014a}. The modification used by Duque \textit{et al}.
was originally proposed in Chiari \textit{et al}. \citep{Chiari2013},
and allows the separation of the ionisation ICS from the total ``inelastic''
ICS provided by the IAM-SCAR approach. The data of Bull \textit{et
al}. is determined from measurements of ion and electron currents
using the Beer-Lambert law \citep{Bull2014}, and agrees quite well
with the BEB results in terms of both its shape and in the position
of the cross section maximum, while being generally lower in magnitude
compared to the theoretical results. As the sole experimental data
available, and from a group with a long history of making reliable
ionisation cross section measurements, we use the Bull \textit{et
al}. ionisation ICS as the basis for our proposed ionisation cross
section. For energies above the maximum considered by Bull \textit{et
al}. (285 eV), we make use of the modified IAM-SCAR results of Duque
\textit{et al}. For energies below the minimum considered by Bull
\textit{et al}. (10 eV), we specify an ionisation threshold of 9.69
eV, as was obtained by Dampc \textit{et al}. \citep{Dampc2017} through
the analysis of THFA photoelectron spectra. This resulting initial
proposed ionisation cross section can also be found plotted in Figure
\ref{fig:Ionisation-cross-section}.

\subsection{Electron attachment cross section}

\begin{figure}
\begin{centering}
\includegraphics[scale=0.5]{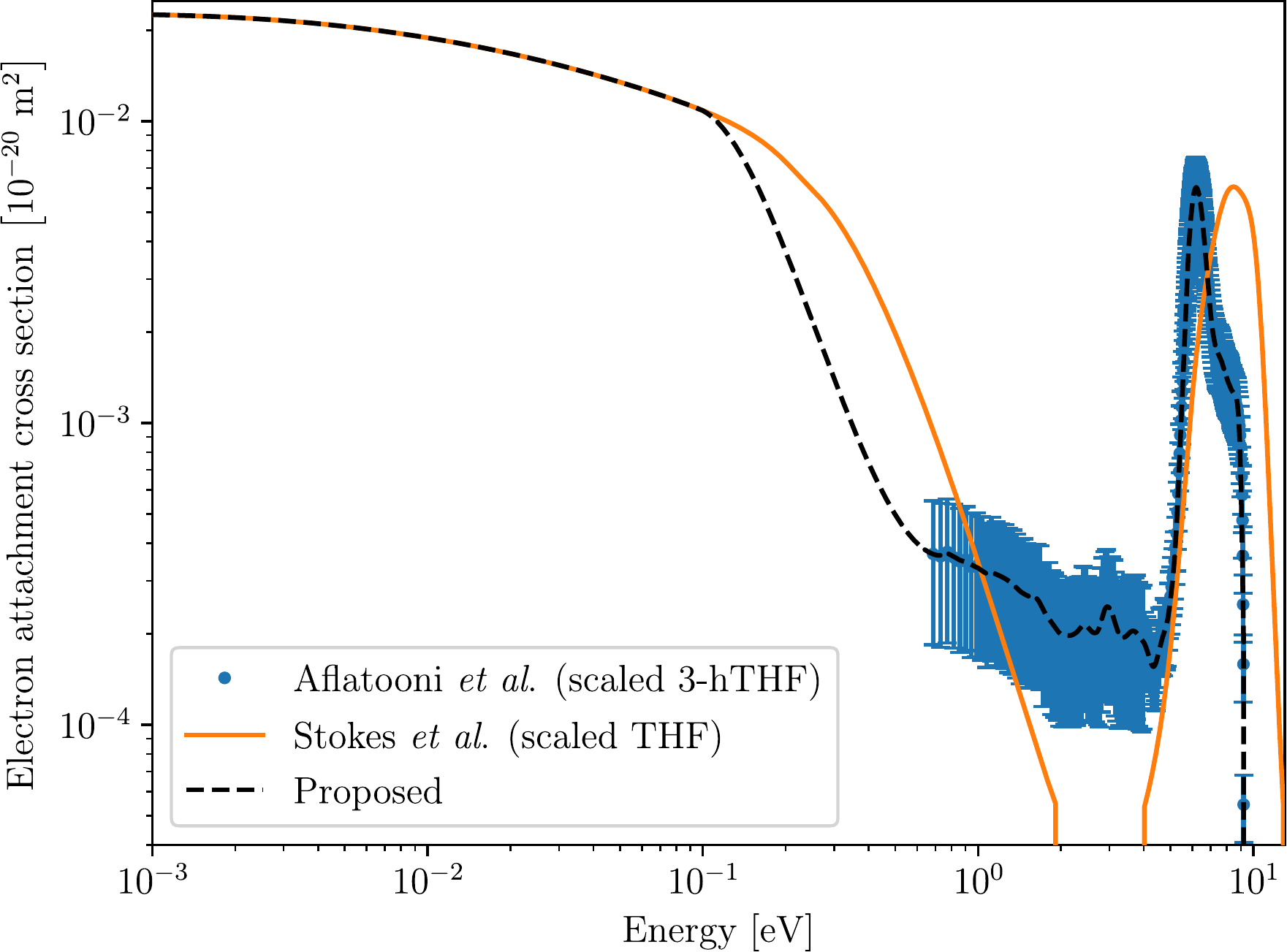}
\par\end{centering}
\caption{\label{fig:Attachment-cross-section}Proposed electron attachment
cross section, alongside the previous results \citep{Aflatooni2006,Stokes2020}
from which it was derived. See also legend in figure.}
\end{figure}
To our knowledge, there are currently no attachment scattering data
available for THFA in the literature. There are, however, dissociative
electron attachment (DEA) cross sections available for some other
structurally similar biomolecules, including THF \citep{Aflatooni2006,Garland2013,Janeckova2014,Casey2017,DeUrquijo2019a,Stokes2020}
and 3-hydroxytetrahydrofuran (3-hTHF) \citep{Aflatooni2006}, from
which we can obtain a very rough initial estimate for that of THFA.
Brunger \citep{Brunger2017} noted that the most important electronic-structure
quantities for determining the relative magnitude of DEA cross sections
between biomolecules is the dipole moment and dipole polarisability
of each molecule, with the dipole polarisability being the most significant.
While of course this is very much a first order approximation, it
does provide at least some physical basis to what follows. On average,
across all five of its conformers at room temperature \citep{Limao-Vieira2014},
THFA has a large dipole polarisability of $\sim63.38a_{0}^{3}$ \citep{Duque2014a,Limao-Vieira2014},
where $a_{0}$ is the Bohr radius. For construction of our proposed
electron attachment cross section for THFA, we prioritise using the
available 3-hTHF DEA data over that for THF, as both 3-hTHF and THFA
contain a hydroxyl group, the presence of which has been shown experimentally
to enhance DEA \citep{Aflatooni2006}. Specifically, we choose to
use the 3-hTHF DEA measurement of Aflatooni \textit{et al}. \citep{Aflatooni2006}.
Compared to THFA, 3-hTHF has an average dipole polarisability of $\sim50.8281a_{0}^{3}$
\citep{Zecca2008,Vizcaino2008}, a value we obtain simply by averaging
the theoretically-determined values of its two most energetically
stable conformers of $50.6779a_{0}^{3}$ and $50.9782a_{0}^{3}$ \citep{Zecca2008}.
Accordingly, we scale up the Aflatooni \textit{et al}. 3-hTHF DEA
measurement by the ratio between the THFA and 3-hTHF dipole polarisabilities
($\times1.247$). This then completes the DEA component of our proposed
attachment cross section. At energies below 0.1 eV, where DEA is impossible
and any attachment is inherently non-dissociative, we make use of
the electron attachment cross section for THF derived by Stokes\textit{
et al}. \citep{Stokes2020} using a neural network-based analysis
of swarm transport data. Of course it is also necessary to scale this
cross section ($\times52.71$), so that the peak magnitude for DEA
matches that for the scaled 3-hTHF DEA measurement used at higher
energies. Finally, we perform a smoothing cubic spline interpolation
\citep{Reinsch1967} over the entire range being considered. The resulting
proposed electron attachment cross section for THFA can be found plotted
in Figure \ref{fig:Attachment-cross-section}, alongside the scaled
THF and 3-hTHF counterparts it was derived from.

\subsection{Grand total cross section}

\begin{figure}
\begin{centering}
\includegraphics[scale=0.5]{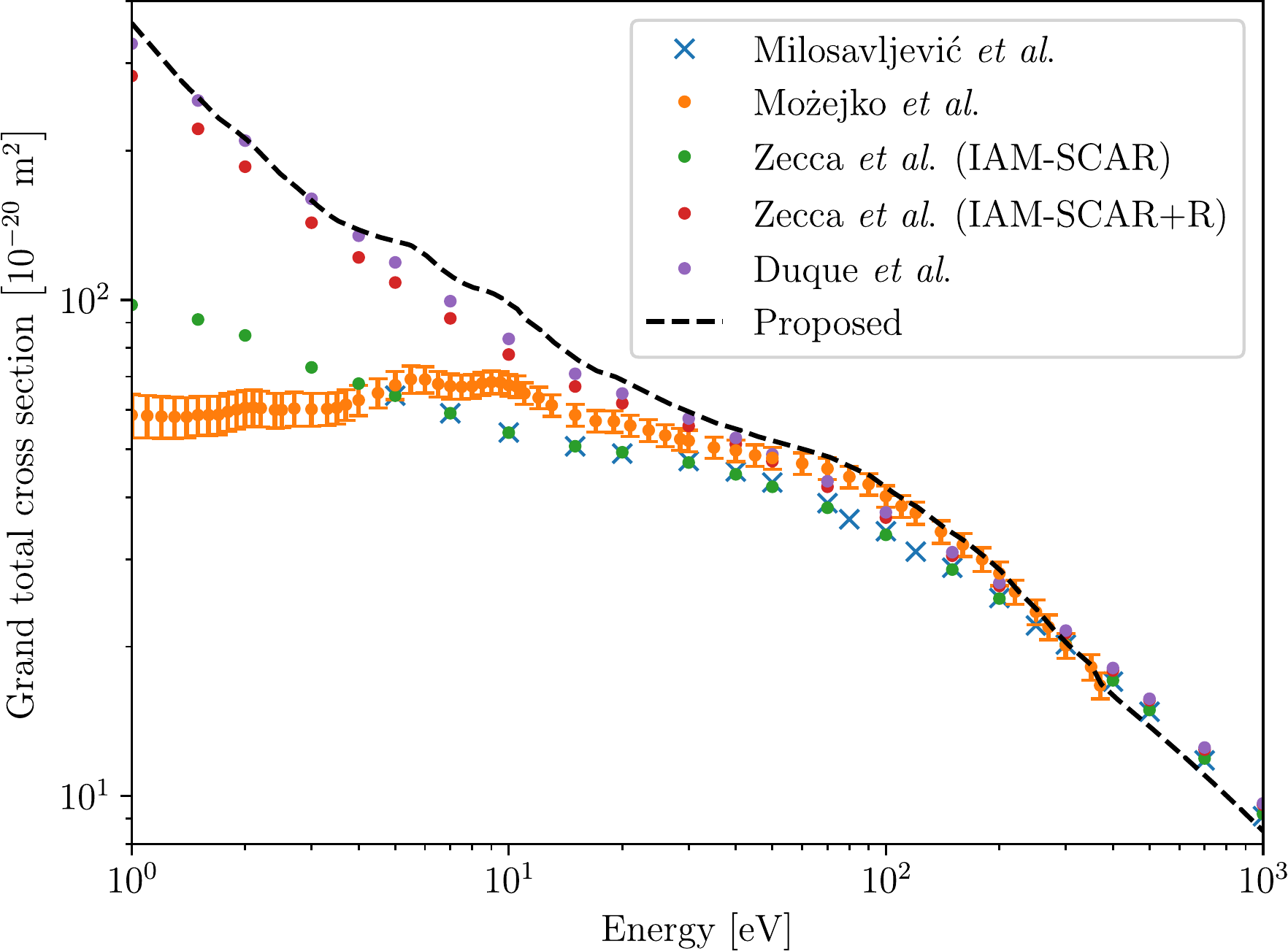}
\par\end{centering}
\caption{\label{fig:Total-cross-section}Proposed grand total cross section,
alongside previous experimental and theoretical total cross sections
\citep{Milosavljevic2006,Mozejko2006,Zecca2011,Duque2014a}. See also
legend in figure.}
\end{figure}
Grand total cross section (TCS) scattering data for electron-THFA
includes the theoretical TCS results from Milosavljević \textit{et
al}. \citep{Milosavljevic2006}, the experimental TCS data of Możejko
\textit{et al}. \citep{Mozejko2006}, the theoretical TCS data of
Zecca \textit{et al}. \citep{Zecca2011}, and the theoretical TCS
data of Duque \textit{et al}. \citep{Duque2014a}. These are summarised
in Figure \ref{fig:Total-cross-section}. The experimental data of
Możejko \textit{et al}. are derived by applying the Beer-Lambert formula
\citep{Mozejko2006}, to the attenuated and unattenuated electron
beam intensities, in a linear transmission experiment. Those data
are not corrected for the forward-scattering effect \citep{Mozejko2006},
and are therefore expected to be missing the significant effect of
the rotational cross sections at lower energies. The theoretical data
of Milosavljević \textit{et al}. is calculated using the IAM-SCAR
approach, which yields combined elastic, electronic excitation, neutral
dissociation and ionisation processes, while lacking contributions
from rotational, vibrational excitation and DEA processes. The TCS
data of Zecca \textit{et al}. is calculated using the IAM-SCAR+R procedure,
and includes all processes except for vibrational excitation and DEA.
Here rotational excitation is included using a Born approximation-based
method. The data of Duque \textit{et al}. is also calculated using
the IAM-SCAR+R procedure, but the innovation of Chiari \textit{et
al}. \citep{Chiari2013} is also employed in this case to separate
the ionisation channel from the rest of the ``inelastic'' data.
The TCS data of Duque \textit{et al}. therefore also includes all
processes except for vibrational excitation and DEA, but is able to
individually resolve the elastic, rotational, discrete electronic-state
excitation and ionisation cross sections. For our proposed grand TCS
for THFA, we prioritise the experimental data of Możejko \textit{et
al}. over the theoretical results. However, to use this data we must
first correct for the forward-scattering effect by increasing the
magnitude of this cross section at lower energies. To determine the
extent of this correction, we use the TCS data of Duque \textit{et
al}. as a guide, to which we add our proposed vibrational excitation
and DEA cross sections to form an initial approximate grand TCS. Next,
we scale the experimental data of Możejko \textit{et al}. so as to
best match this approximation. In particular, we scale by an energy-dependent
correction of the power-law form $1+5.169\varepsilon^{-1.050}$, which
has the greatest effect at smaller energies, while leaving the experimental
data at higher energies unaffected. At energies above 370 eV, we use
the same approximate grand TCS data derived from Duque \textit{et
al}., but scaled down ($\times0.88$) so as to improve continuity
with the experimental data of Możejko \textit{et al}. The resulting
proposed grand TCS is plotted in Figure \ref{fig:Total-cross-section},
alongside the various results from the literature \citep{Milosavljevic2006,Mozejko2006,Zecca2011,Duque2014a}.

\subsection{Rotational cross section}

\begin{figure}
\begin{centering}
\includegraphics[scale=0.5]{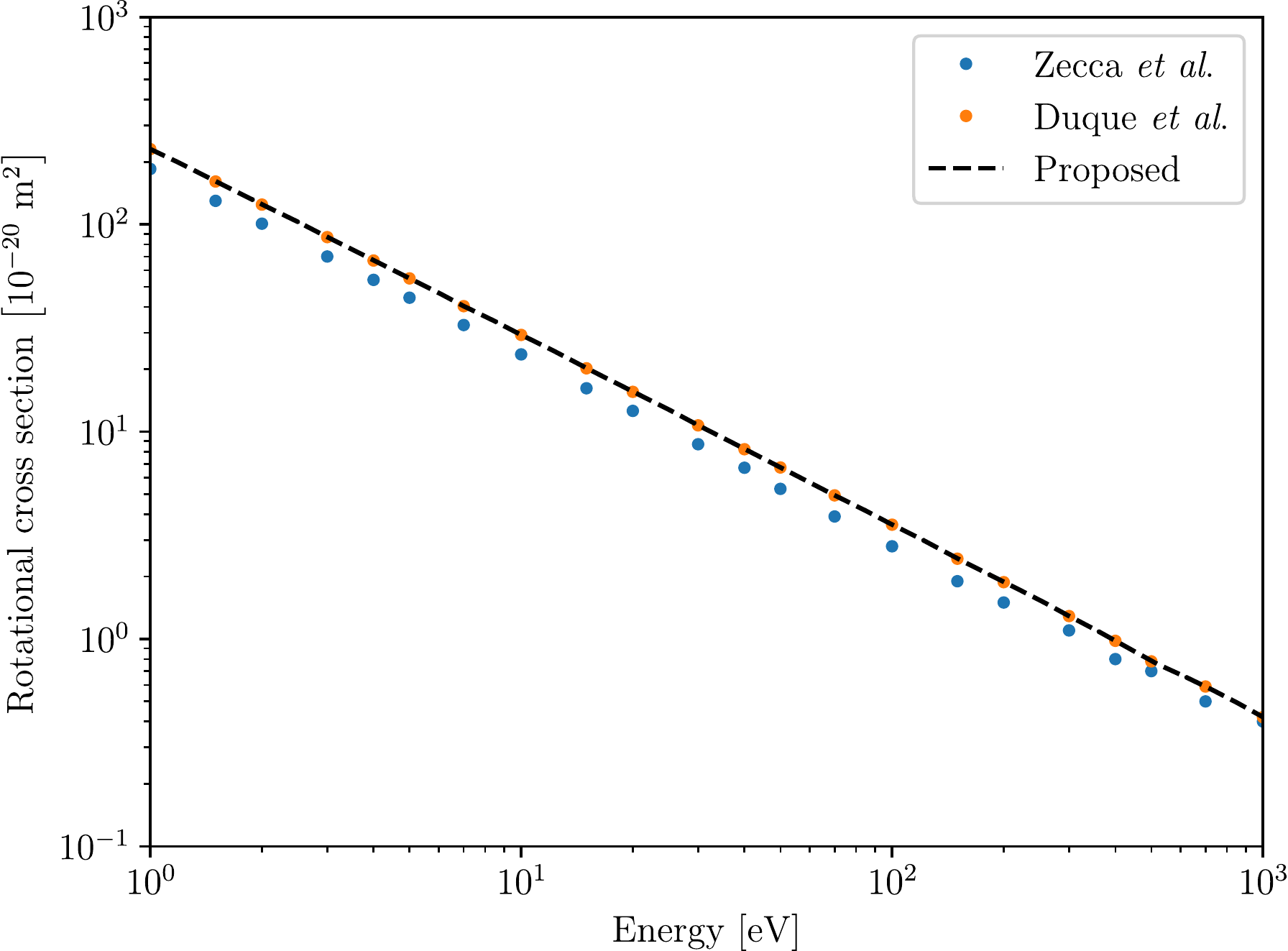}
\par\end{centering}
\caption{\label{fig:Rotational-cross-section}Proposed rotational integral
cross section, alongside previous theoretical results \citep{Zecca2011,Duque2014a}.
See also legend in figure.}
\end{figure}
Rotational scattering data for electron-THFA collisions include the
theoretical ICS data of Zecca \textit{et al}. \citep{Zecca2011},
and the theoretical ICS data of Duque \textit{et al}. \citep{Duque2014a}.
These are plotted in Figure \ref{fig:Rotational-cross-section}. Both
of these rotational cross sections are derived using the first Born
approximation (FBA), with Zecca \textit{et al}. calculating the rotational
excitation cross section, for $J\rightarrow J^{\prime}$ in THFA at
300 K, by weighting the population for the $J^{\mathrm{th}}$ rotational
quantum number at that temperature and estimating the average excitation
energy from the corresponding rotational constants. On the other hand,
Duque \textit{et al}. used the procedure of Jain \citep{Jain1988}.
All the integral rotational cross section data is given in terms of
a single summed rotational cross section, with the ICS data of Duque
\textit{et al}. calculated from assuming an average rotational threshold
energy of 0.74 meV. As we are relying on the TCS data of Duque \textit{et
al}. to guide the form of our proposed grand TCS, for consistency
we also make use here of the data from Duque \textit{et al}. for our
proposed rotational ICS, as shown in Figure \ref{fig:Rotational-cross-section}.

\subsection{\label{subsec:Elastic-cross-section}Elastic cross section}

\begin{figure}
\begin{centering}
\includegraphics[scale=0.5]{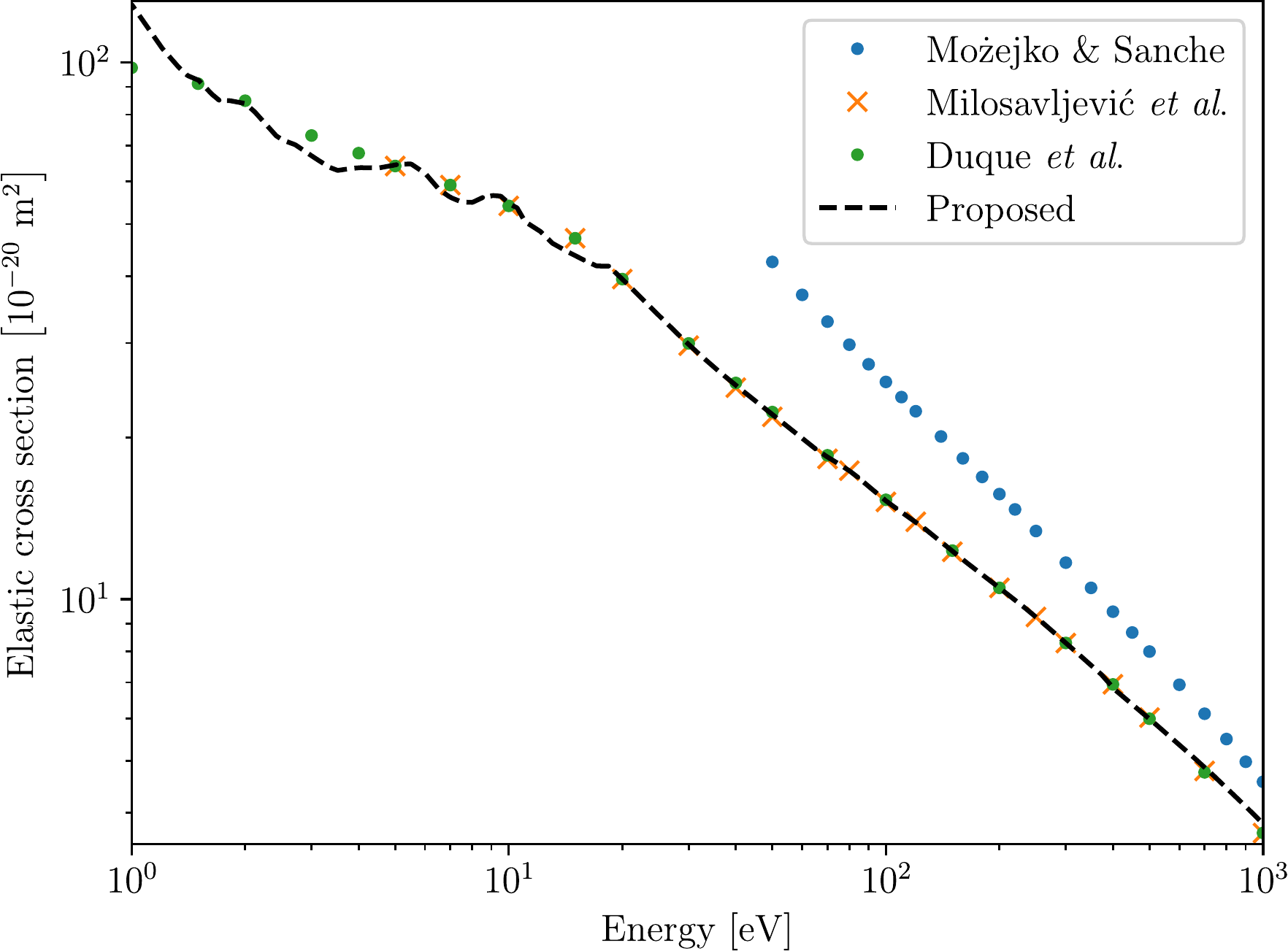}
\par\end{centering}
\caption{\label{fig:Elastic-cross-section}Proposed elastic integral cross
section, alongside previous theoretical results \citep{Mozejko2005,Milosavljevic2006,Duque2014a}.
See also legend in figure.}
\end{figure}
Elastic scattering data for electron-THFA collisions include the theoretical
DCS and ICS data of Możejko and Sanche \citep{Mozejko2005}, the theoretical
ICS data of Milosavljević \textit{et al}. \citep{Milosavljevic2006}
and the theoretical ICS data of Duque \textit{et al}. \citep{Duque2014a}.
These are summarised in Figure \ref{fig:Elastic-cross-section}. The
calculations of Możejko and Sanche are performed using the independent-atom
method (IAM), applied with the additivity rule (AR), and have been
superseded by the calculations of Milosavljević \textit{et al}. and
Duque \textit{et al}., which use the IAM-SCAR procedure. That is,
the IAM in conjunction with a screening corrected additivity rule
(SCAR). The ICS data of both Milosavljević \textit{et al}. and Duque
\textit{et al}. are in good agreement with one another, and as such
we base our proposed THFA elastic ICS on both of these results, with
some minor modifications in place for consistency with our proposed
grand TCS. The resulting proposed elastic ICS is plotted in Figure
\ref{fig:Elastic-cross-section}. Even by accounting for all of the
cross sections proposed thus far (outlined in Sections \ref{subsec:Electronic-excitation-cross}–\ref{subsec:Elastic-cross-section}),
there still remains a discrepancy in the grand TCS at intermediate
energies which we attribute to neutral dissociation, as discussed
in the following section.

\subsection{Neutral dissociation cross section}

\begin{figure}
\begin{centering}
\includegraphics[scale=0.5]{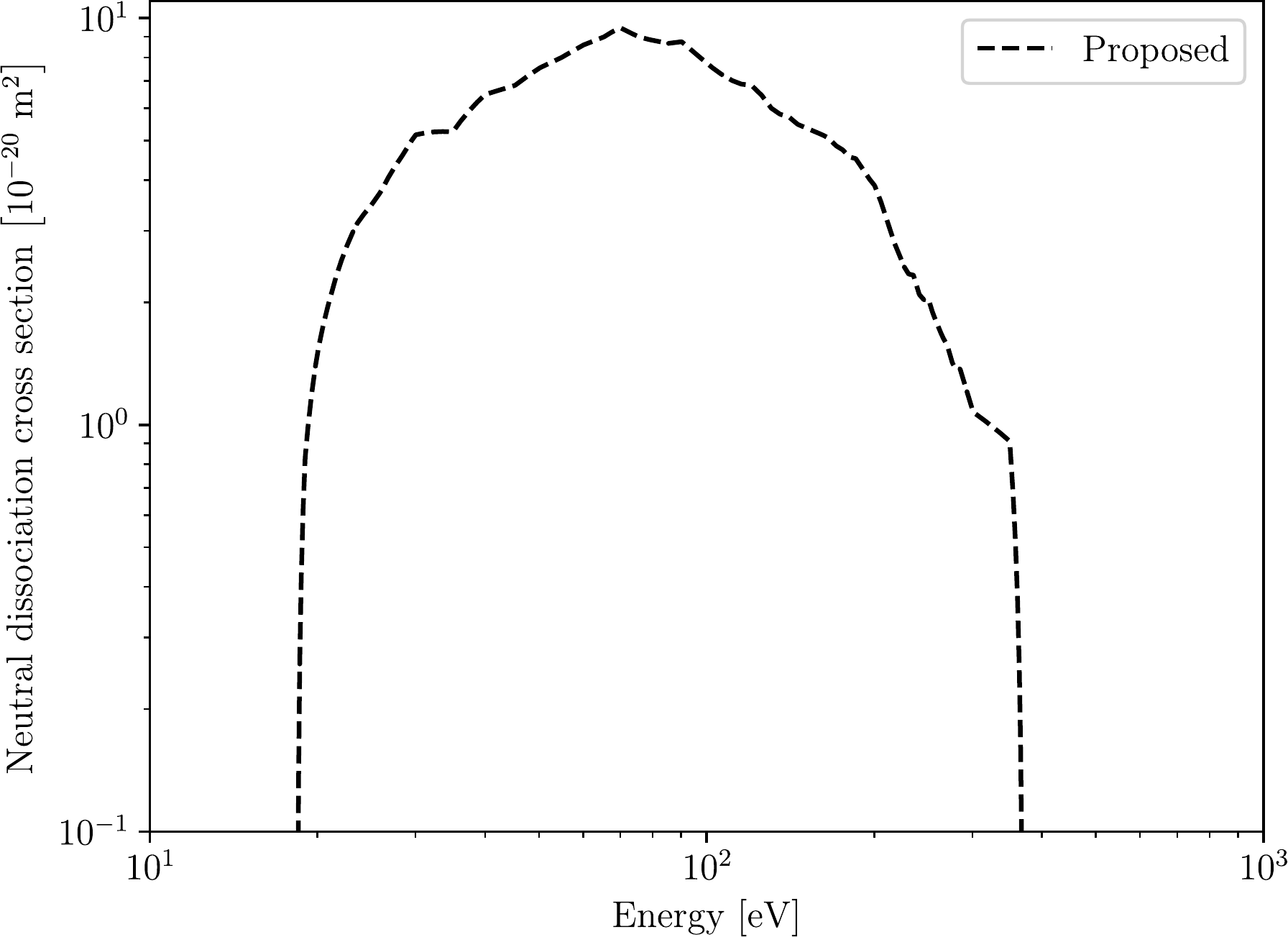}
\par\end{centering}
\caption{\label{fig:Neutral-dissociation-cross-section}Proposed neutral dissociation
cross section. See text for further details.}
\end{figure}
A challenge in obtaining complete sets of electron-biomolecule cross
sections is the intractability of determining the neutral dissociation
integral cross section from scattering experiments directly \citep{Buckman2013}.
Similarly, theoretical results are rare \citep{Brunger2002}. Of course,
in principle, it can be found indirectly by subtracting all the other
scattering cross sections from the TCS, resulting in a remnant that
is attributed to neutral dissociation. Proceeding with this approach
results in the proposed THFA neutral dissociation cross section plotted
in Figure \ref{fig:Neutral-dissociation-cross-section}, with an apparent
threshold energy of $18.4\ \mathrm{eV}$. However, as the accuracy
of this remnant is predicated on the collective accuracy of all other
cross sections outlined thus far, it is not anticipated to be particularly
reliable. As an alternative to this approach, swarm experiments provide
an implicit way of elucidating the neutral dissociation cross section
through assessment of the self-consistency of a cross section set.
Indeed, the neutral dissociation cross section of THF has previously
been characterised through the swarm analysis of Casey \textit{et
al}. \citep{Casey2017} and de Urquijo \textit{et al}. \citep{DeUrquijo2019a},
as well as recently by Stokes \textit{et al}. \citep{Stokes2020}
using the same machine learning formalism described in Section \ref{sec:Neural-network-for}
and employed later in Section \ref{sec:Refined-set-of} of the present
investigation.

\subsection{Quasielastic momentum transfer cross section}

\begin{figure}
\begin{centering}
\includegraphics[scale=0.5]{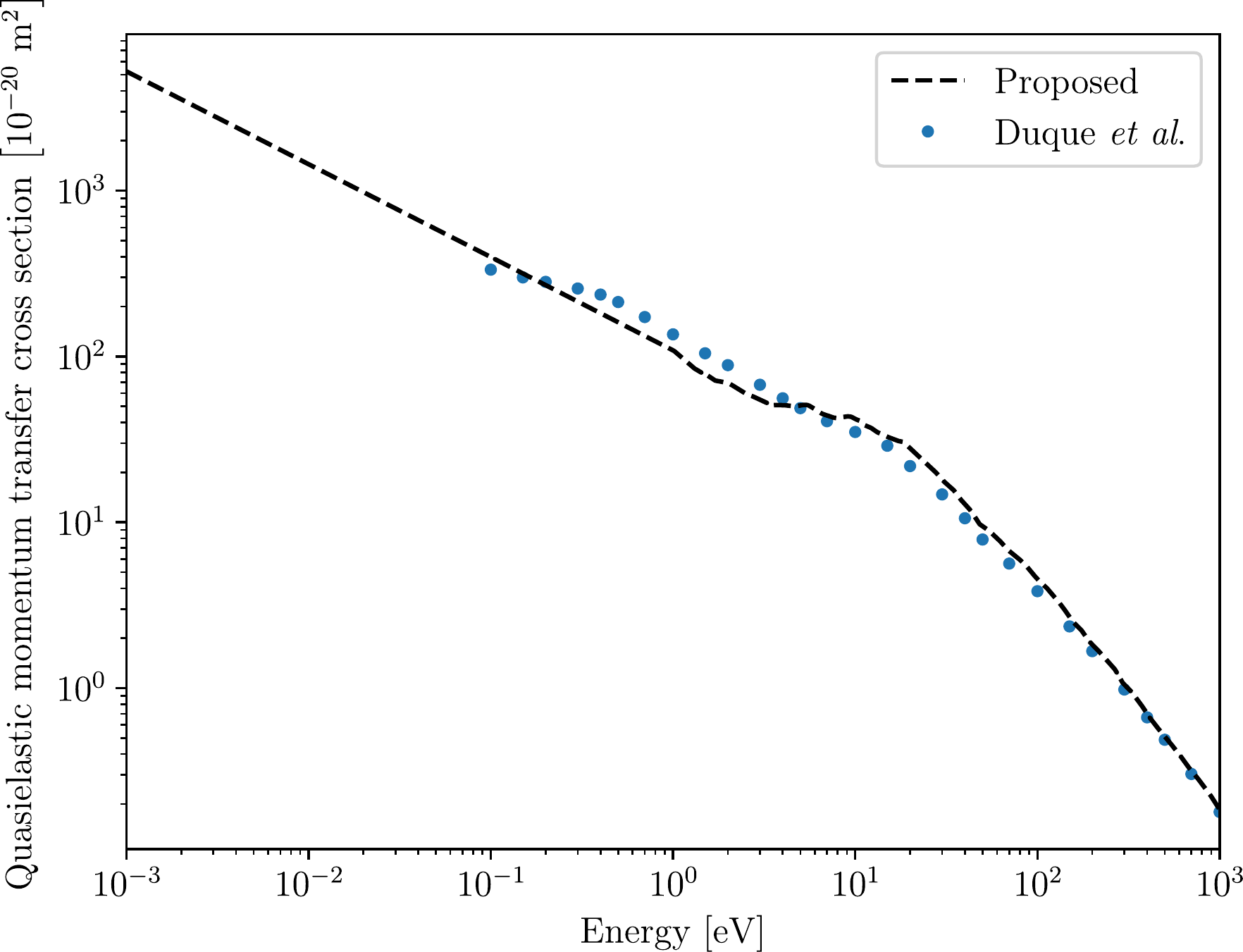}
\par\end{centering}
\caption{\label{fig:Quasielastic-cross-section}Proposed quasielastic momentum
transfer cross section. See text for further details.}
\end{figure}
For the purposes of performing transport calculations, we form a proposed
quasielastic (elastic+rotational) momentum transfer cross section
(MTCS) by scaling our proposed quasielastic ICS by the ratio of quasielastic
MTCS to ICS of Duque \textit{et al}. \citep{Duque2014a}. At energies
below 1 eV, we utilise a power law extrapolation that is fitted to
the quasielastic MTCS of Duque \textit{et al}. in this regime. Our
resulting proposed quasielastic MTCS is plotted in Figure \ref{fig:Quasielastic-cross-section}.

\section{\label{sec:Transport-coefficients-of-PROPOSED}Pulsed-Townsend swarm
measurements for assessing the self-consistency of the proposed cross
section set}

\begin{figure}
\begin{centering}
\includegraphics[scale=0.5]{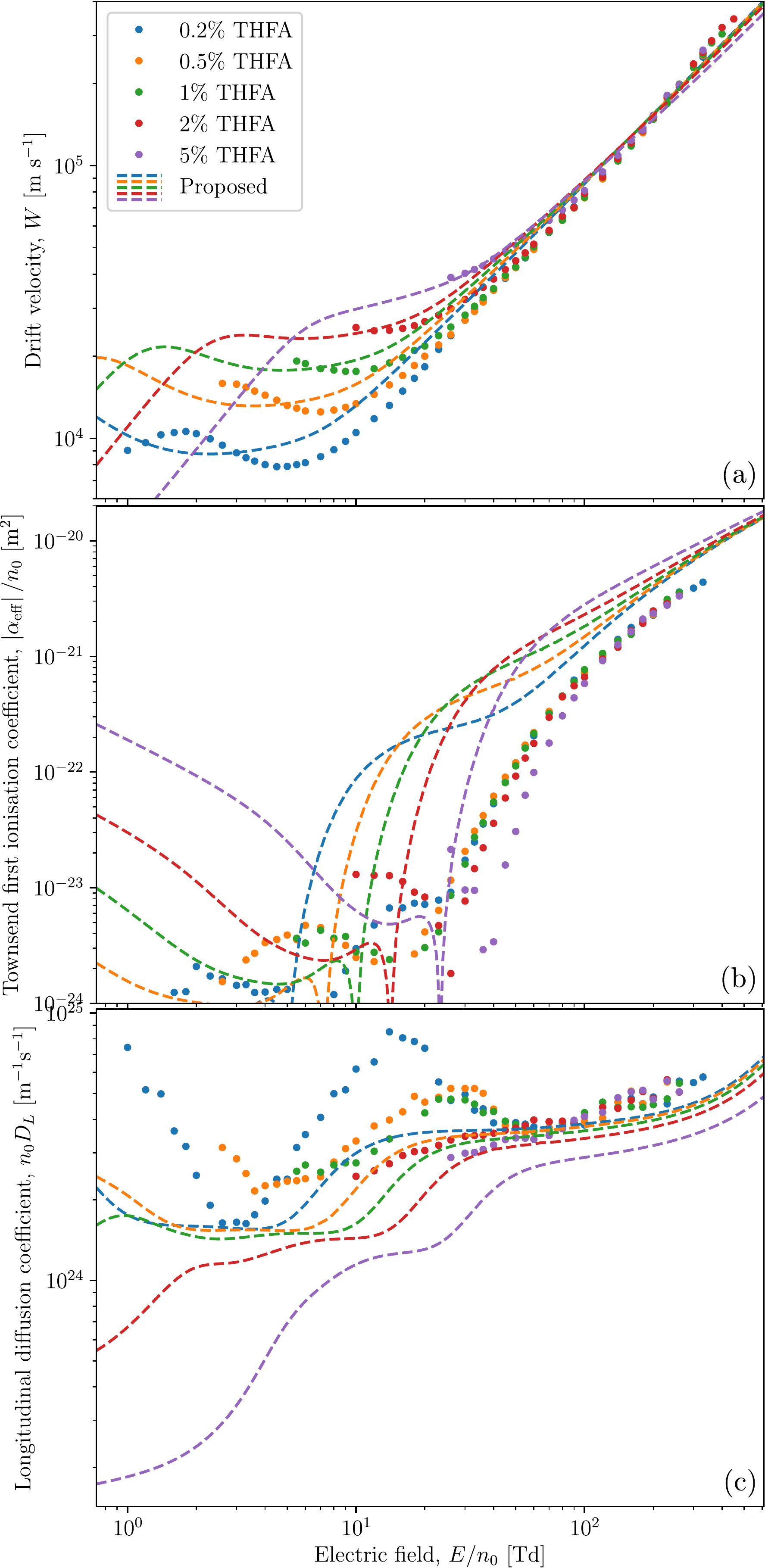}
\par\end{centering}
\caption{\label{fig:Proposed-transport-coeffici}Measured pulsed-Townsend swarm
transport coefficients (markers) of (a), the drift velocity, $W$,
(b), the effective Townsend first ionisation coefficients, $\alpha_{\mathrm{eff}}/n_{0}$,
and (c), the longitudinal diffusion coefficients, $n_{0}D_{L}$. Simulated
transport coefficients (dashed curves), derived from the proposed
cross section set presented in Section \ref{sec:Electron-THFA-cross-section},
are also plotted for comparison.}
\end{figure}
To assess the quality of our proposed set of electron-THFA cross sections,
we perform a number of pulsed-Townsend swarm experiments from which
we obtain drift velocities, effective Townsend first ionisation coefficients,
and longitudinal diffusion coefficients for admixtures of THFA in
argon. We consider mixture ratios of 0.2\%, 0.5\%, 1\%, 2\%, and 5\%
THFA, across a range of reduced electric fields, from 1 Td to 450
Td, where $1\ \mathrm{Td}=1\ \mathrm{Townsend}=10^{-21}\ \mathrm{V\ m^{2}}$.
Results from these measurements — which are tabulated in our appendix
Tables \ref{tab:Measured-pulsed-Townsend-0.2}–\ref{tab:Measured-pulsed-Townsend-5}
— were derived from a pulsed-Townsend apparatus whose technique and
methods of analysis of the electron avalanche waveforms have been
accounted for in detail previously \citep{Hernndez-vila2002,Basurto2013}.
The pulsed Townsend method is based on the measurement of the total
displacement current due to the motion of electrons and ions within
a parallel plate capacitor that produces a highly homogeneous field
upon the application of a highly stable and very low ripple DC voltage
between 150 V to 5 kV, depending on the $E/n_{0}$ and $n_{0}$ conditions
of the experiment. The capacitor through which the charge carriers
drift and react consists of an aluminium cathode and a non-magnetic
stainless steel anode of 12 cm diameter each, separated by an accurately
measured distance of 3.1 cm to within an accuracy of 0.025 mm. The
initial photoelectrons are generated from the cathode by the incidence
of a UV laser pulse (1–2 mJ, 355 nm, 3 ns). When the dominant processes
involved in the avalanche are due to the electrons and stable, non-reacting
ions, then any collisional ionisation or attachment events are due
only to electrons. Furthermore, since the electron drift velocity
is $10^{2}$–$10^{3}$ times larger than that of the ions, the resulting
total current can readily be separated into a fast component due mostly
to the electrons, followed by that of the slower ions. The analysis
of the electron component leads to the derivation of the flux electron
drift velocity, $W$, and the density-normalised effective ionisation
coefficient, $\alpha_{\mathrm{eff}}/n_{0}=\left(\alpha-\eta\right)/n_{0}$,
where $\alpha$ and $\eta$ are the ionisation and electron attachment
coefficients. A very stable voltage in the range 0.2–5 kV was applied
to the anode in order to produce the highly homogeneous electric field
$E$, according to the $E/n_{0}$ value selected and the gas density
$n_{0}$ in the discharge vessel. The stated purity of the commercial
THFA sample used was 99.0\% (Sigma-Aldrich) and that of Ar was 99.995\%
(Praxair). Because of the very low vapour pressure of THFA, namely
0.2 Torr at 293 K, the maximum pressure allowed in the discharge vessel
was 0.18 Torr. This very low filling pressure value hindered the measurements
of the electron swarm coefficients for pure THFA. The mere presence
of the electron avalanche produces a space charge field which is superposed
to the externally applied one. Thus care must be taken to keep the
minimum external voltage high enough so that the space charge field
is smaller than 1\% of that applied to the anode. With the present
configuration, with an interelectrode distance of 3.1 cm, we could
only measure THFA mixtures with Ar successfully from 0.2\% to 5\%
THFA. The minimum external voltage was 200 V. The measurements were
performed at room temperature in the range 293–300 K, measured with
a precision of $\pm0.5\ \mathrm{K}$, while the gas mixture pressure
was monitored with an absolute pressure capacitance gauge ($\pm0.15\%$
uncertainty). The displacement current due to the electrons was measured
with a very low-noise, 40 MHz amplifier with a transimpedance of $10^{5}$
V/A. The measured electron transients were analysed using the formula
for the electron current in the external circuit derived by Brambring
\citep{Brambring1964}:
\begin{equation}
I\left(t\right)=\frac{n_{0}qW}{2L}e^{\alpha_{\mathrm{eff}}Wt}\left\{ \mathrm{erfc}\left[\frac{\left(W+\alpha_{\mathrm{eff}}D_{L}\right)t-L}{\sqrt{4D_{L}t}}\right]-e^{\frac{W+\alpha_{\mathrm{eff}}D_{L}}{D_{L}}L}\mathrm{erfc}\left[\frac{\left(W+\alpha_{\mathrm{eff}}D_{L}\right)t+L}{\sqrt{4D_{L}t}}\right]\right\} ,
\end{equation}
where $L$ is the drift distance and $\mathrm{erfc}\left(x\right)=1-\frac{2}{\sqrt{\pi}}\int_{0}^{x}e^{-u^{2}}\mathrm{d}u$
is the complementary error function. Thus we have three swarm parameters
to determine, namely, $W$, $\alpha_{\mathrm{eff}}$ and $D_{L}$.
The process is simplified by determining initial values of $W$ and
$\alpha_{\mathrm{eff}}$, derived from a basic, geometrical analysis
\citep{Bekstein2012,Urquijo-Carmona1980}, and inserted into a simulator
to fit the whole transient, thereby obtaining $D_{L}$ and refined
values of $\alpha_{\mathrm{eff}}$ and $W$. Typical uncertainties
for $W$, $\alpha_{\mathrm{eff}}$ and $D_{L}$ were $\pm2\%$, $\pm6\%$
and $\pm12\%$, respectively. Note that these transport coefficients
can be related to the net ionisation frequency, $R_{\mathrm{net}}$,
the bulk drift velocity, $W_{B}$, and the bulk diffusion coefficient,
$D_{B,L}$, via \citep{Casey2020}:
\begin{eqnarray}
R_{\mathrm{net}} & = & \alpha_{\mathrm{eff}}W,\\
W_{B} & = & W+\alpha_{\mathrm{eff}}D_{L},\\
D_{B,L} & = & D_{L}.
\end{eqnarray}
Using our proposed cross section set, we apply a well-benchmarked
multi-term solution of Boltzmann's equation \citep{White2018,Boyle2017,White2003}
to derive simulated pulsed-Townsend transport coefficients for comparison
to our admixture measurements. For calculating these admixture transport
coefficients, we use the argon cross section set present in the Puech
database \citep{Puech} on LXCat \citep{Pancheshnyi2012,Pitchford2017,LXCat}.
The simulated transport coefficients are plotted alongside the experimental
values in Figure \ref{fig:Proposed-transport-coeffici}. Figure \ref{fig:Proposed-transport-coeffici}(a)
compares the drift velocities and shows qualitative agreement at low-to-intermediate
fields, with the proposed data set underestimating its magnitude slightly
at the highest fields considered. Interestingly, while both measured
and simulated drift velocities exhibit negative differential conductivity
(NDC) — drift velocity decreasing with increasing reduced electric
field — there is disagreement in the extent of NDC, with the simulated
drift velocities exhibiting NDC over a much larger range of fields.
Figure \ref{fig:Proposed-transport-coeffici}(b) compares the effective
Townsend first ionisation coefficients and shows that those which
result from the proposed set overestimate the magnitude in the electropositive
regime, and underestimate the magnitude in the electronegative regime.
That is, overall, the simulated effective Townsend first ionisation
coefficients are too positive, suggesting that an increase in the
magnitude of the proposed attachment cross section is required, as
well as a decrease in magnitude of the proposed ionisation cross section.
Figure \ref{fig:Proposed-transport-coeffici}(c) compares the longitudinal
diffusion coefficients, and shows the poorest agreement between simulation
and experiment across all of the considered transport coefficients
with the simulated diffusion coefficients consistently underestimating
the measurements. Fortunately, the general shape of the simulated
diffusion coefficients appears to be in fair qualitative agreement
with experiment.

\section{\label{sec:Refined-set-of}Refined electron-THFA cross sections}

Given the results in Figure \ref{fig:Proposed-transport-coeffici},
we now employ the neural network model, Eq. \eqref{eq:nn}, to solve
the inverse problem of mapping from our admixture swarm measurements
to a selection of desired electron-THFA cross sections. Specifically,
we choose to fit the neutral dissociation cross section, electron
attachment cross section, electron impact ionisation cross section
and quasielastic MTCS. By replacing those cross sections in our earlier
proposed set, with those predicted by the neural network, we complete
our refined set of electron-THFA cross sections.

\subsection{Refined neutral dissociation cross section}

\begin{figure}
\begin{centering}
\includegraphics[scale=0.5]{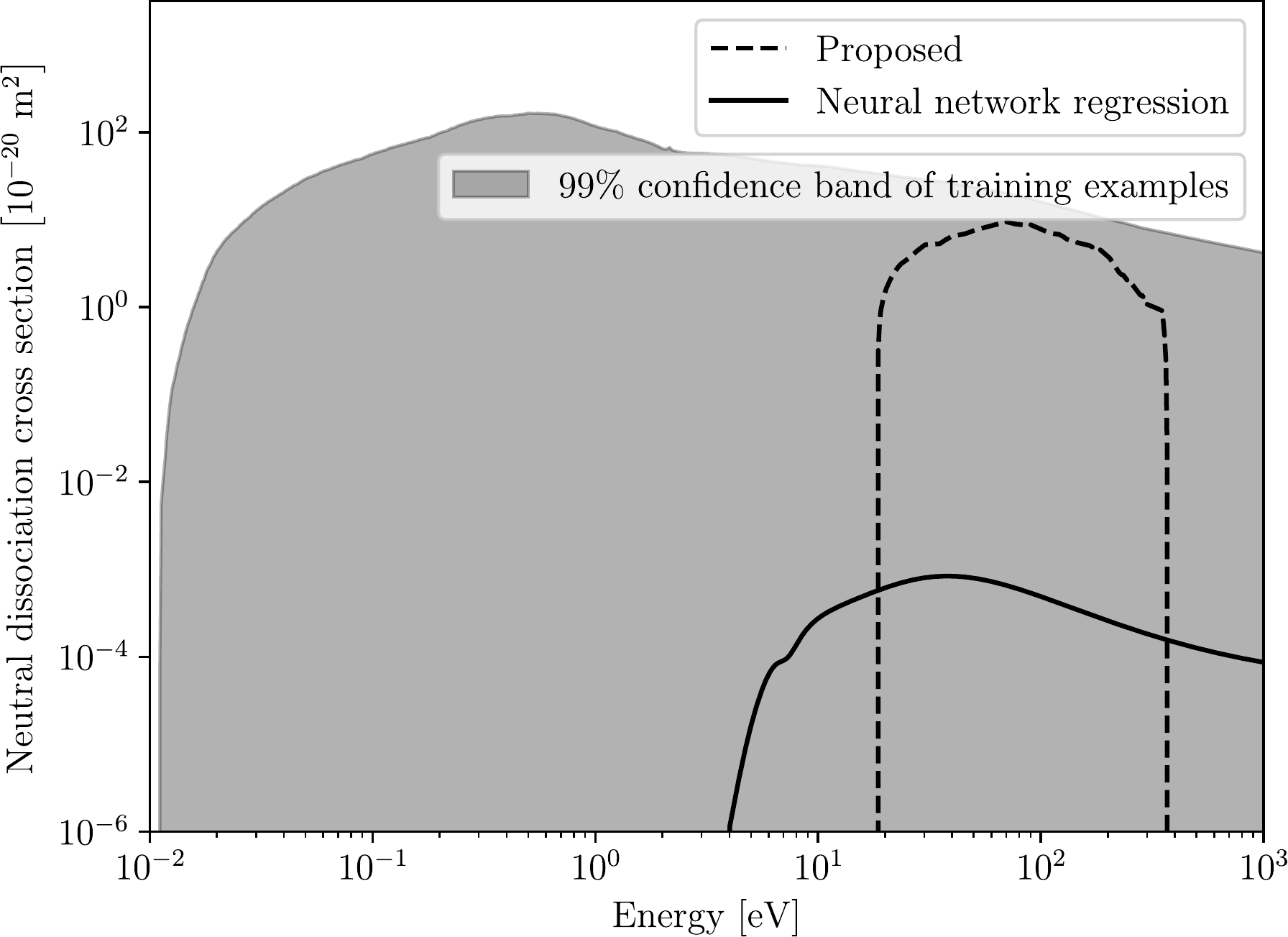}
\par\end{centering}
\caption{\label{fig:Neutral-dissociation-refined}Neural network regression
results for the THFA neutral dissociation cross section, alongside
the earlier proposed ICS for comparison. See also legend in figure.}
\end{figure}
As very little can be stated about the nature of neutral dissociation
in THFA, we choose to place no explicit constraints on the neutral
dissociation cross section when training the neural network, thus
forming training data, using Eq. \eqref{eq:mixture}, to directly
combine random pairs of excitation cross sections from the LXCat project.
The resulting confidence band of training examples is plotted in Figure
\ref{fig:Neutral-dissociation-refined}, alongside the subsequent
refined fit provided by the neural network, as well as the original
proposed cross section for comparison. Compared to that which was
proposed originally, the neural network predicts a neutral dissociation
cross section that is substantially smaller in magnitude, peaking
at $8.4\times10^{-24}\ \mathrm{m}^{2}$ versus $9.4\times10^{-20}\ \mathrm{m}^{2}$,
while also having a smaller threshold energy, with a value of 3.96
eV versus 18.4 eV. Promisingly, the neural network has also resolved
a plausible high-energy tail — a feature that is lost when determining
neural dissociation as a residual of the grand TCS.

\subsection{Refined electron attachment cross section}

\begin{figure}
\begin{centering}
\includegraphics[scale=0.5]{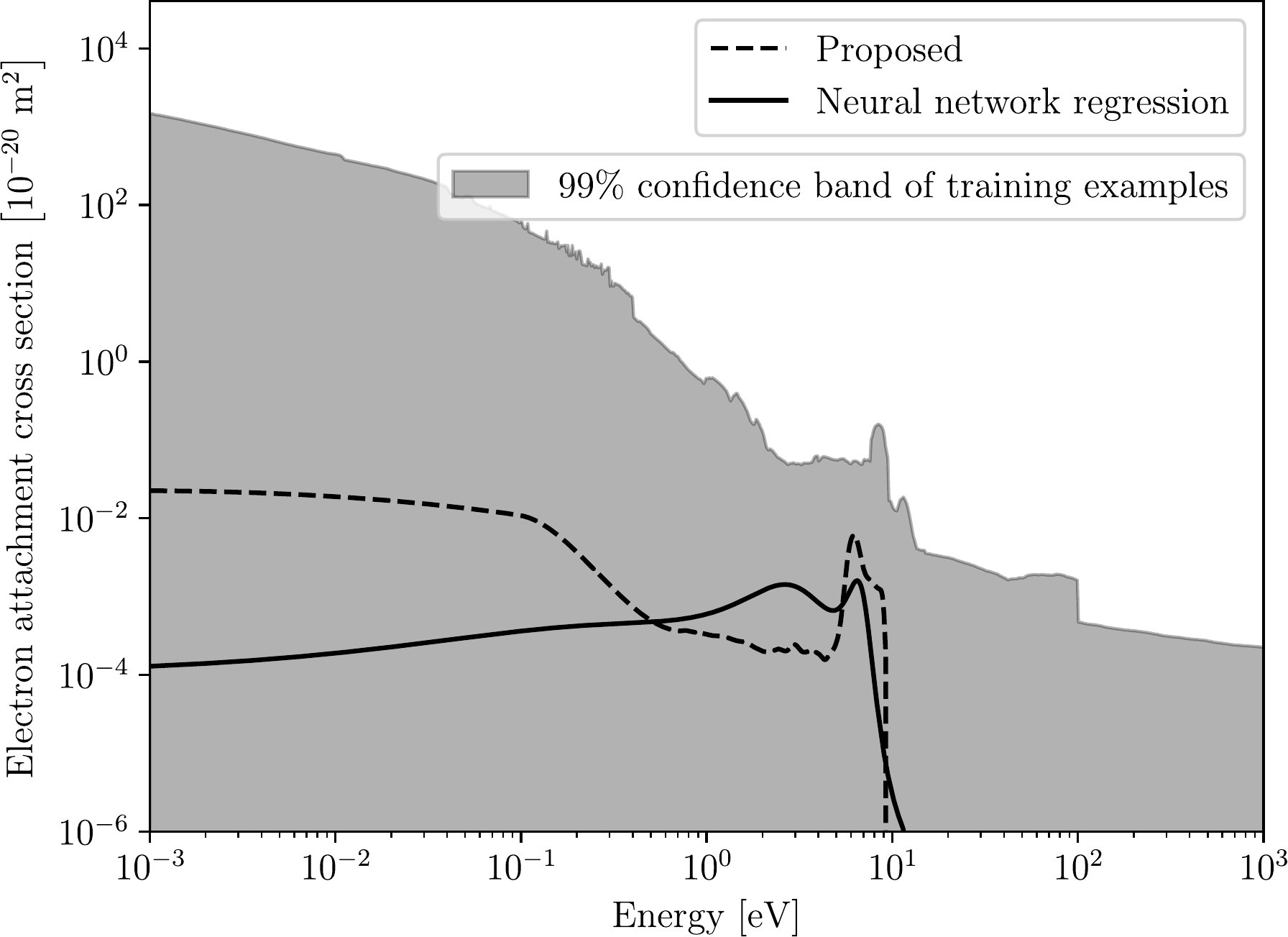}
\par\end{centering}
\caption{\label{fig:Attachment-refined}Neural network regression results for
the THFA dissociative electron attachment cross section, alongside
the earlier proposed ICS for comparison. See also legend in figure.}
\end{figure}
Due to the current lack of electron-THFA attachment data in the literature,
the attachment cross sections used for training the neural network
are chosen in a similarly unconstrained fashion as those used to refine
neutral dissociation. That is, Eq. \eqref{eq:mixture} is used to
combine random pairs of LXCat attachment cross sections for training,
while enforcing no additional explicit cross section constraints.
The resulting confidence band of training examples is plotted in Figure
\ref{fig:Attachment-refined}, alongside the refined fit provided
by the neural network, as well as the original proposed attachment
cross section. The neural network predicts an attachment cross section
that is more uniform than that which was proposed initially, thus
resulting in some substantial differences in different energy regimes.
For example at low energies, near $10^{-3}\ \mathrm{eV}$, the refinement
is over two orders of magnitude smaller than the initial proposal.
At intermediate energies, around 3 eV, the refinement rises slightly
in magnitude while the proposal drops significantly, resulting in
the refinement exceeding the proposal by almost an order of magnitude.
Both the refinement and the initial proposal have a peak near 6 eV,
although the refined attachment cross section has a peak magnitude
that is almost a factor of 4 smaller than that for the initial proposal.
At higher energies, past this peak, both attachment cross sections
decay fairly rapidly in magnitude, with practically no attachment
beyond 12 eV in the refinement, compared to 10 eV for the original
proposed data.

\subsection{Refined electron impact ionisation cross section}

\begin{figure}
\begin{centering}
\includegraphics[scale=0.5]{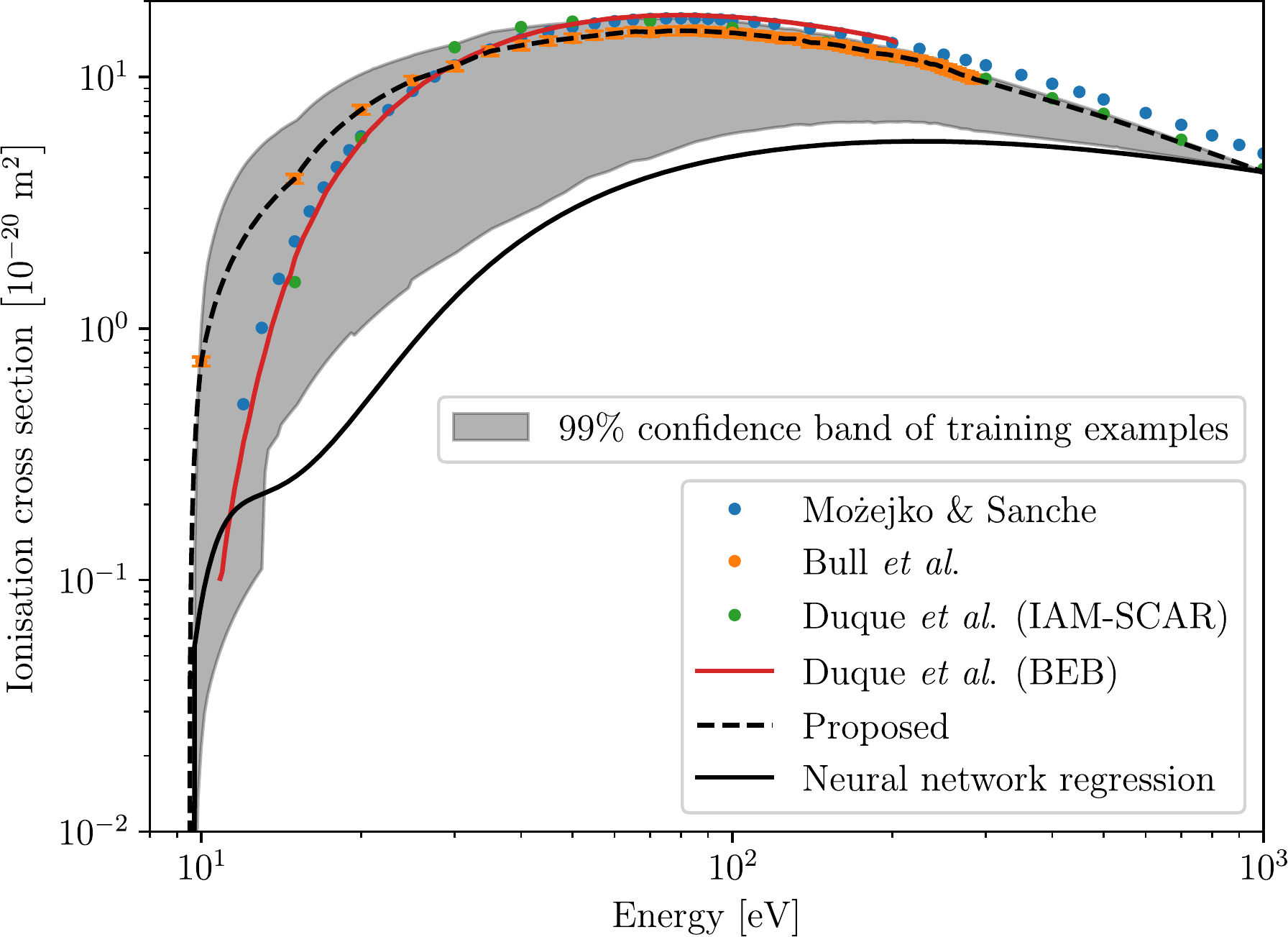}
\par\end{centering}
\caption{\label{fig:Ionisation-refined}Neural network regression results for
the THFA electron impact ionisation cross section, alongside the earlier
proposed ICS and the previous experimental and theoretical results
\citep{Mozejko2005,Bull2014,Duque2014a} from which it was derived.
See also legend in figure.}
\end{figure}
Given the general agreement among the ionisation cross sections reviewed
thus far, we choose to constrain the ionisation training examples
to within the vicinity of our resulting initial proposal. We make
these constraints particularly stringent at higher energies, where
the theoretical results are expected to be more accurate, and where
our swarm analysis is expected to be less informative. Across all
training examples we use the same threshold energy of 9.69 eV, as
was done for our initial proposed ionisation cross section. Consequently,
the ionisation training cross sections are formed using a formula
very similar to Eq. \eqref{eq:mixture}:
\begin{equation}
\sigma\left(\varepsilon\right)=\sigma_{1}^{1-r\left(\varepsilon\right)}\left(\varepsilon\right)\sigma_{2}^{r\left(\varepsilon\right)}\left(\varepsilon+\varepsilon_{2}-9.69\ \mathrm{eV}\right),
\end{equation}
where $\sigma\left(\varepsilon\right)$ is the ionisation cross section
used for training, $\sigma_{1}\left(\varepsilon\right)$ is our initial
proposed ionisation cross section, $\sigma_{2}\left(\varepsilon\right)$
is a randomly chosen LXCat ionisation cross section, $\varepsilon_{2}$
is that cross section's corresponding threshold energy, and we also
introduce an energy-dependent mixing ratio that varies from $0.2$
to $0.0$ as the energy varies from the ionisation threshold to $10^{3}\ \mathrm{eV}$:
\begin{equation}
r\left(\varepsilon\right)=0.2\frac{\ln\left(\frac{\varepsilon}{10^{3}\ \mathrm{eV}}\right)}{\ln\left(\frac{9.69\ \mathrm{eV}}{10^{3}\ \mathrm{eV}}\right)}.
\end{equation}
The resulting constrained confidence band of training examples is
plotted in Figure \ref{fig:Ionisation-refined}, alongside the refined
fit provided by the neural network, as well as the original proposed
ionisation cross section. As was expected, the refined ionisation
cross section predicted by the neural network is smaller in magnitude,
peaking at $5.5\times10^{-20}\ \mathrm{m}^{2}$ compared to $15\times10^{-20}\ \mathrm{m}^{2}$
for what we proposed initially. Nonetheless, such a large drop in
magnitude ($\sim2.7$ times) is a little concerning given the reputation
of Bull \textit{et al}. \citep{Bull2014} group. However, THFA is
a very difficult molecule to work with experimentally, so such a mismatch
may indeed be possible in this case.

\subsection{Refined quasielastic momentum transfer cross section}

\begin{figure}
\begin{centering}
\includegraphics[scale=0.5]{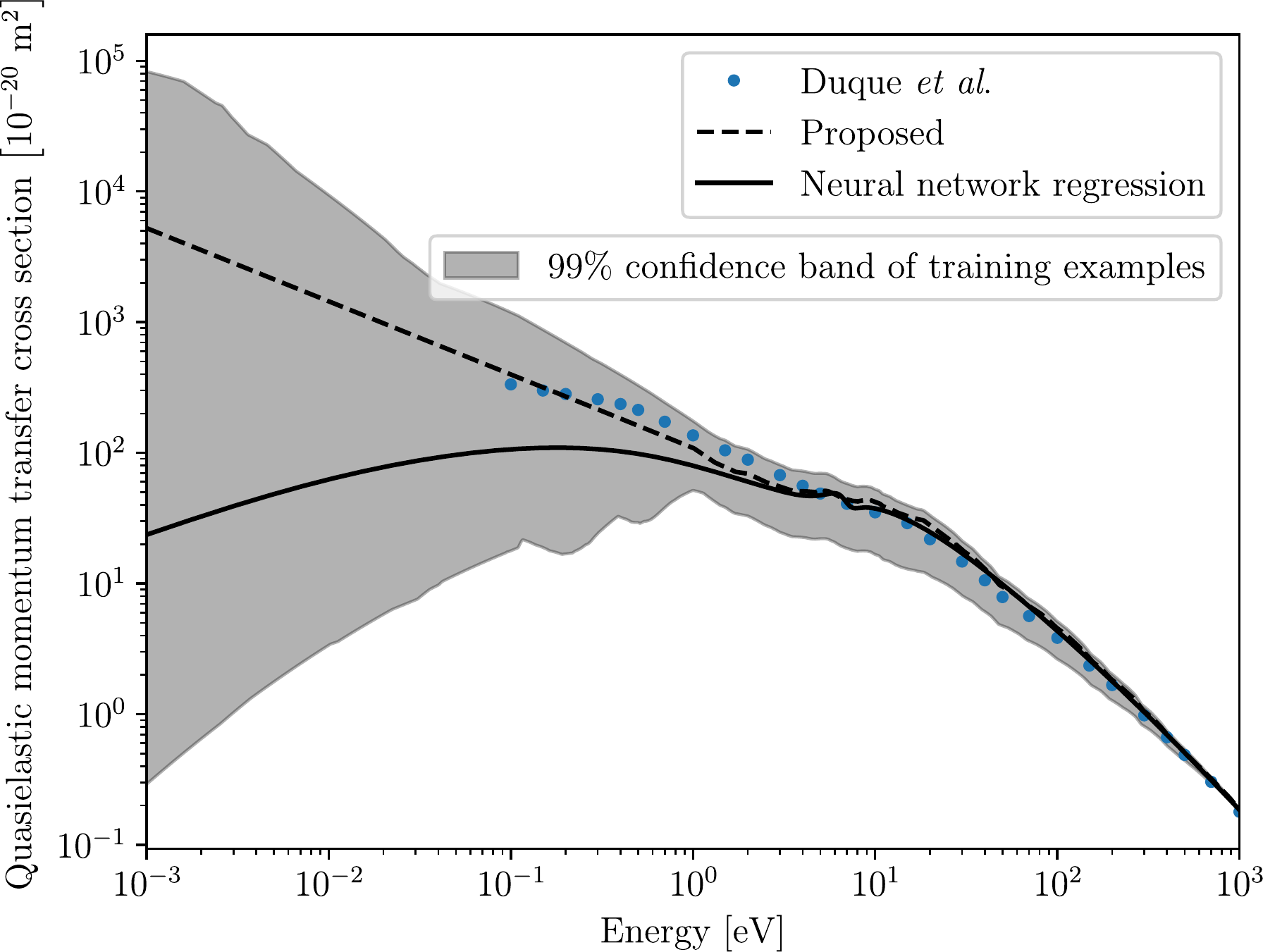}
\par\end{centering}
\caption{\label{fig:Quasielastic-refined}Neural network regression results
for the THFA quasielastic MTCS, alongside the earlier proposed MTCS
for comparison. See also legend in figure.}
\end{figure}
As with the ionisation cross section, we expect our proposed quasielastic
MTCS to be most accurate at higher energies. As such, we proceed similarly
to our approach with ionisation and sample each quasielastic MTCS
for training using the following formula:
\begin{equation}
\sigma\left(\varepsilon\right)=\sigma_{1}^{1-r\left(\varepsilon\right)}\left(\varepsilon\right)\sigma_{2}^{r\left(\varepsilon\right)}\left(\varepsilon\right),
\end{equation}
where $\sigma\left(\varepsilon\right)$ is the MTCS cross section
used for training, $\sigma_{1}\left(\varepsilon\right)$ is our initial
proposed quasielastic MTCS, $\sigma_{2}\left(\varepsilon\right)$
is a randomly chosen LXCat elastic cross section, and we define here
the energy-dependent mixing ratio that varies from $1.0$ to $0.15$
to $0.0$ as the energy varies from $10^{-3}\ \mathrm{eV}$ to $1\ \mathrm{eV}$
to $10^{3}\ \mathrm{eV}$:
\begin{equation}
r\left(\varepsilon\right)=0.15\left\{ \begin{array}{cc}
1-\frac{17}{3}\frac{\ln\left(\frac{\varepsilon}{1\ \mathrm{eV}}\right)}{\ln\left(\frac{10^{3}\ \mathrm{eV}}{1\ \mathrm{eV}}\right)}, & 10^{-3}\ \mathrm{eV}\leq\varepsilon\leq1\ \mathrm{eV},\\
1-\frac{\ln\left(\frac{\varepsilon}{1\ \mathrm{eV}}\right)}{\ln\left(\frac{10^{3}\ \mathrm{eV}}{1\ \mathrm{eV}}\right)}, & 1\ \mathrm{eV}\leq\varepsilon\leq10^{3}\ \mathrm{eV}.
\end{array}\right.
\end{equation}
The resulting constrained confidence band of training examples is
plotted in Figure \ref{fig:Quasielastic-refined}, alongside the refined
fit provided by the neural network, as well as the original proposed
quasielastic MTCS. The refined quasielastic MTCS provided by the neural
network is essentially identical to the proposal at energies above
1 eV. Below 1 eV, however, the neural network predicts quasielastic
MTCS that is significantly smaller than the proposal, steadily decreasing
in relative magnitude as energy is decreased. The greatest difference
occurs at $10^{-3}\ \mathrm{eV}$, where the refined quasielastic
MTCS is roughly two orders of magnitude smaller than its counterpart
from our initial proposed set.

\subsection{\label{subsec:Refined-grand-total}Refined grand total cross section}

\begin{figure}
\begin{centering}
\includegraphics[scale=0.5]{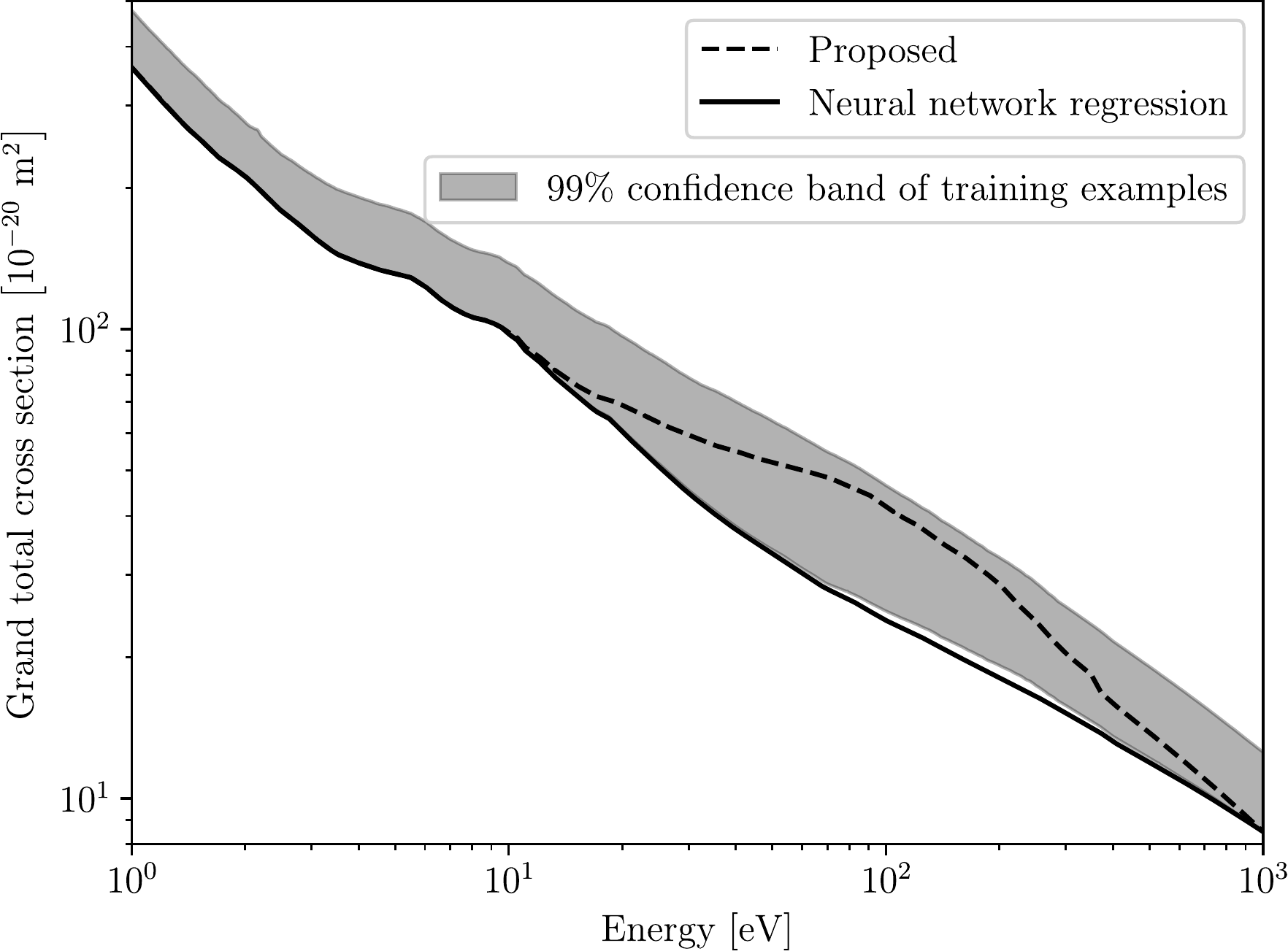}
\par\end{centering}
\caption{\label{fig:Total-refined}Neural network regression results for the
THFA grand TCS, alongside the original proposed TCS for comparison.
See also legend in figure.}
\end{figure}
Although it is not considered explicitly, the grand TCS of each cross
section set used for training is naturally affected by the aforementioned
constraints placed on the neutral dissociation, attachment, and ionisation
cross sections. The resulting constrained confidence band of training
examples is plotted in Figure \ref{fig:Total-refined}, alongside
the refined fit provided by the neural network, as well as the original
proposed grand TCS. The neural network predictions result in a grand
TCS that is smaller than the initial proposal for energies above 10
eV or so, while coinciding below this energy. The greatest difference
arises at 90 eV with a reduction in magnitude of 43\%, most of which
being due to the reduction in the ionisation cross section (see Figure
\ref{fig:Ionisation-refined}).

\section{\label{sec:Transport-coefficients-of-REFINED}Transport coefficients
of the refined electron-THFA cross section set}

\subsection{Refined admixture transport coefficients}

\begin{figure}
\begin{centering}
\includegraphics[scale=0.5]{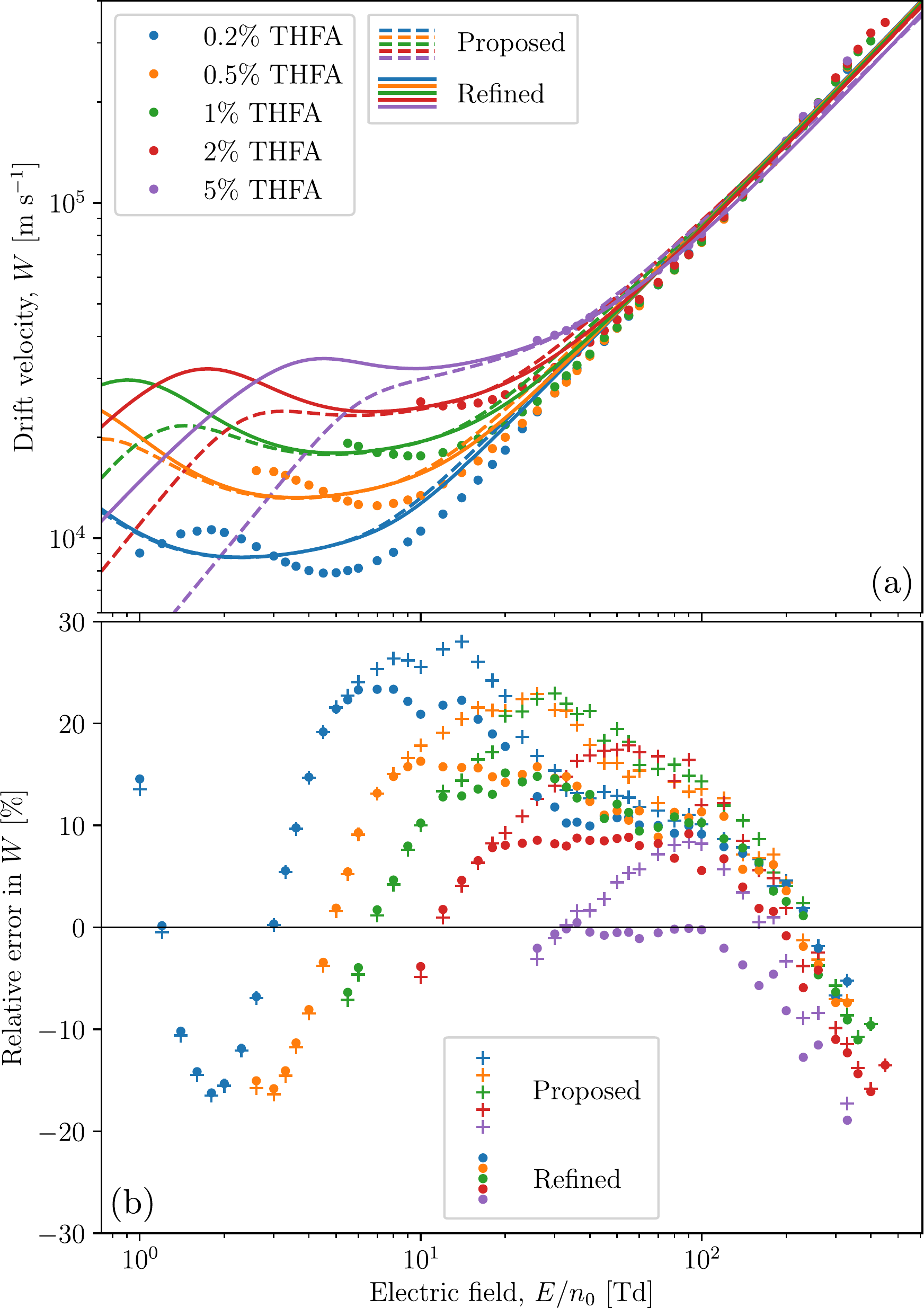}
\par\end{centering}
\caption{\label{fig:Drift-velocity}(a) simulated drift velocities from both
our original proposed data base and our refined data base, compared
to corresponding results from our admixture swarm measurements. (b)
corresponding percentage errors in the simulated values relative to
the swarm measurements. See also legends in figures.}
\end{figure}
\begin{figure}
\begin{centering}
\includegraphics[scale=0.5]{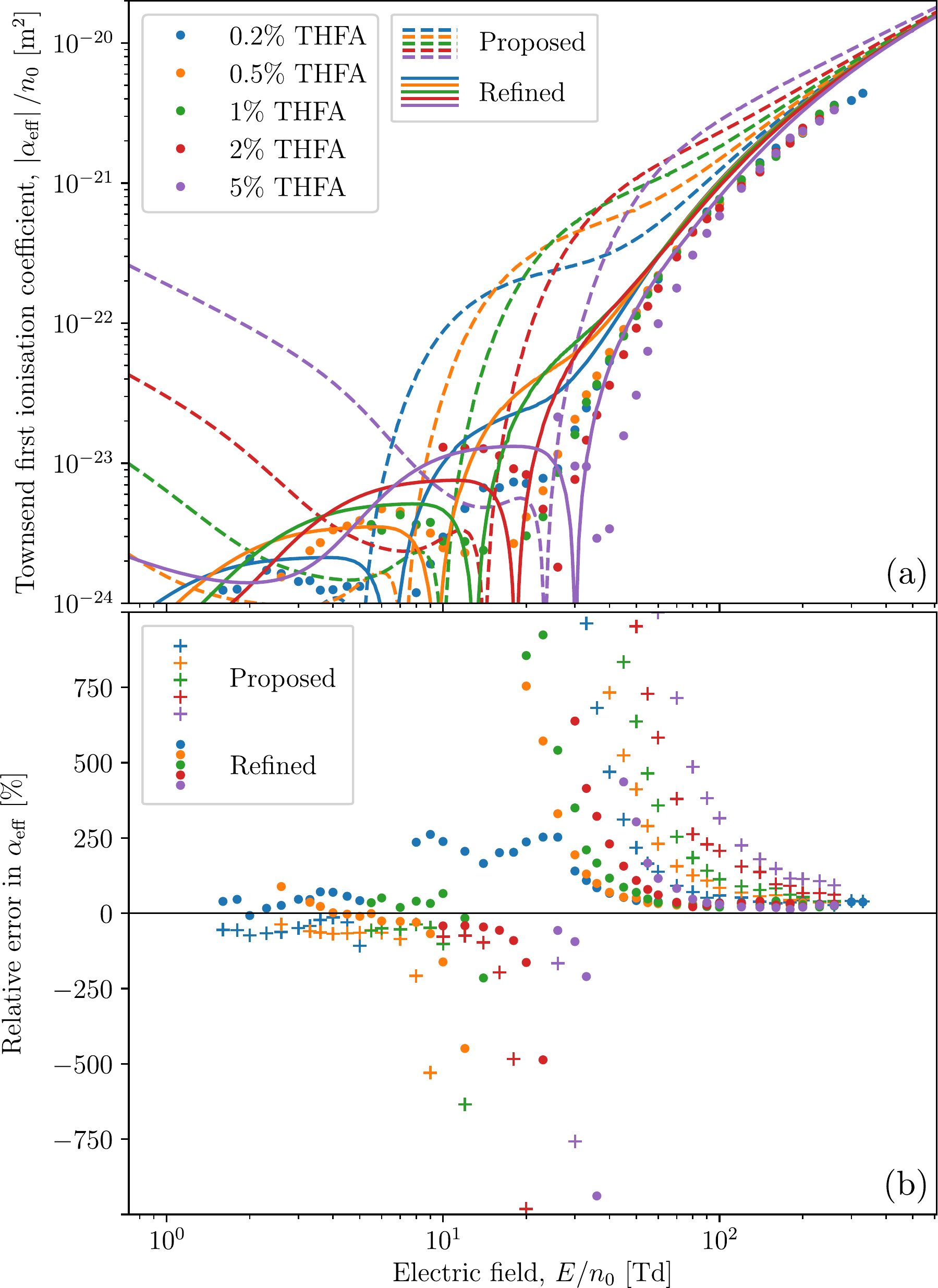}
\par\end{centering}
\caption{\label{fig:Townsend-coefficient}(a) simulated effective Townsend
first ionisation coefficients from both our original proposed data
base and our refined data base, compared to corresponding results
from our admixture swarm measurements. (b) corresponding percentage
errors in the simulated values relative to the swarm measurements.
See also legends in figures.}
\end{figure}
\begin{figure}
\begin{centering}
\includegraphics[scale=0.5]{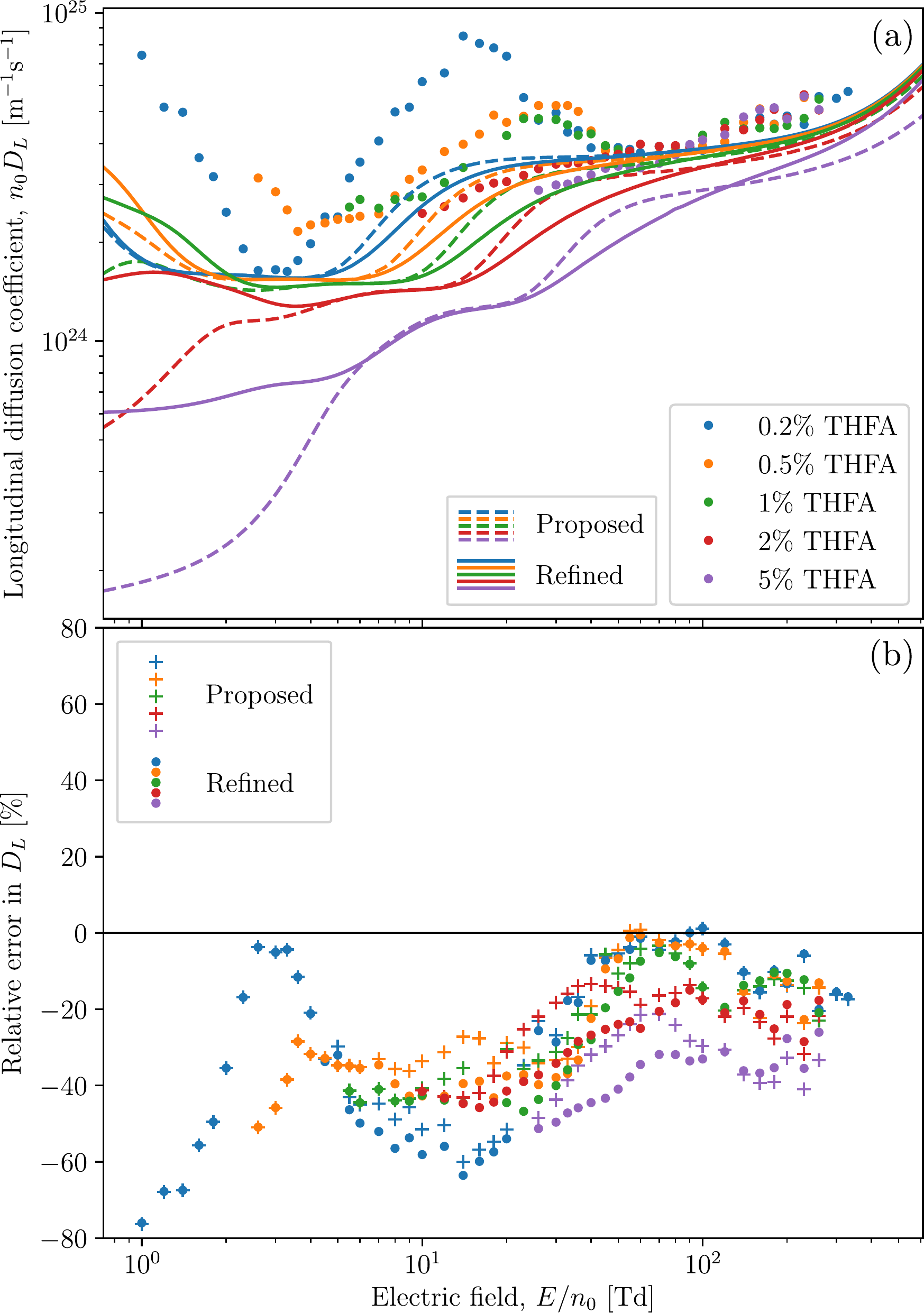}
\par\end{centering}
\caption{\label{fig:Diffusion-coefficient}(a) simulated longitudinal diffusion
coefficients from both our original proposed data base and our refined
data base, compared to corresponding results from our admixture swarm
measurements. (b) corresponding percentage errors in the simulated
values relative to the swarm measurements. See also legends in figures.}
\end{figure}
Using our refined set of electron-THFA cross sections, we plot revised
simulated transport coefficients in Figures \ref{fig:Drift-velocity}–\ref{fig:Diffusion-coefficient}
for comparison to both our admixture swarm measurements, as well as
to the transport coefficients calculated previously for our initial
proposed cross section set. Respectively, the drift velocities, effective
Townsend first ionisation coefficients and longitudinal diffusion
coefficients are each plotted in Figures \ref{fig:Drift-velocity}(a),
\ref{fig:Townsend-coefficient}(a), and \ref{fig:Diffusion-coefficient}(a),
with corresponding percentage error differences plotted in Figures
\ref{fig:Drift-velocity}(b), \ref{fig:Townsend-coefficient}(b),
and \ref{fig:Diffusion-coefficient}(b). Figure \ref{fig:Drift-velocity}
shows that the refined set of electron-THFA cross sections has, in
general, brought the simulated drift velocities closer to the results
from the experimental measurements. There are some instances where
the mismatch has increased, but these are infrequent. The large discrepancies
between simulation and experiment remain at low fields, with the refinement
having very little effect on the drift velocities in this regime.
Figure \ref{fig:Townsend-coefficient} shows a substantial improvement
in the effective Townsend first ionisation coefficients after the
cross-section refinement. Overall, both the relative error and the
shape of the simulated effective Townsend first ionisation coefficients
have improved, with the most benefit seen in the electropositive regime.
Figure \ref{fig:Diffusion-coefficient} shows a slight worsening in
the accuracy of the simulated longitudinal diffusion coefficients
after the neural network refinement, contrary to the other transport
coefficients. It should be noted, however, that the shape of the plotted
longitudinal diffusion coefficients has appeared to improve slightly
with the refined cross section data set.

\subsection{Transport coefficients in pure THFA}

\begin{figure}
\begin{centering}
\includegraphics[scale=0.4]{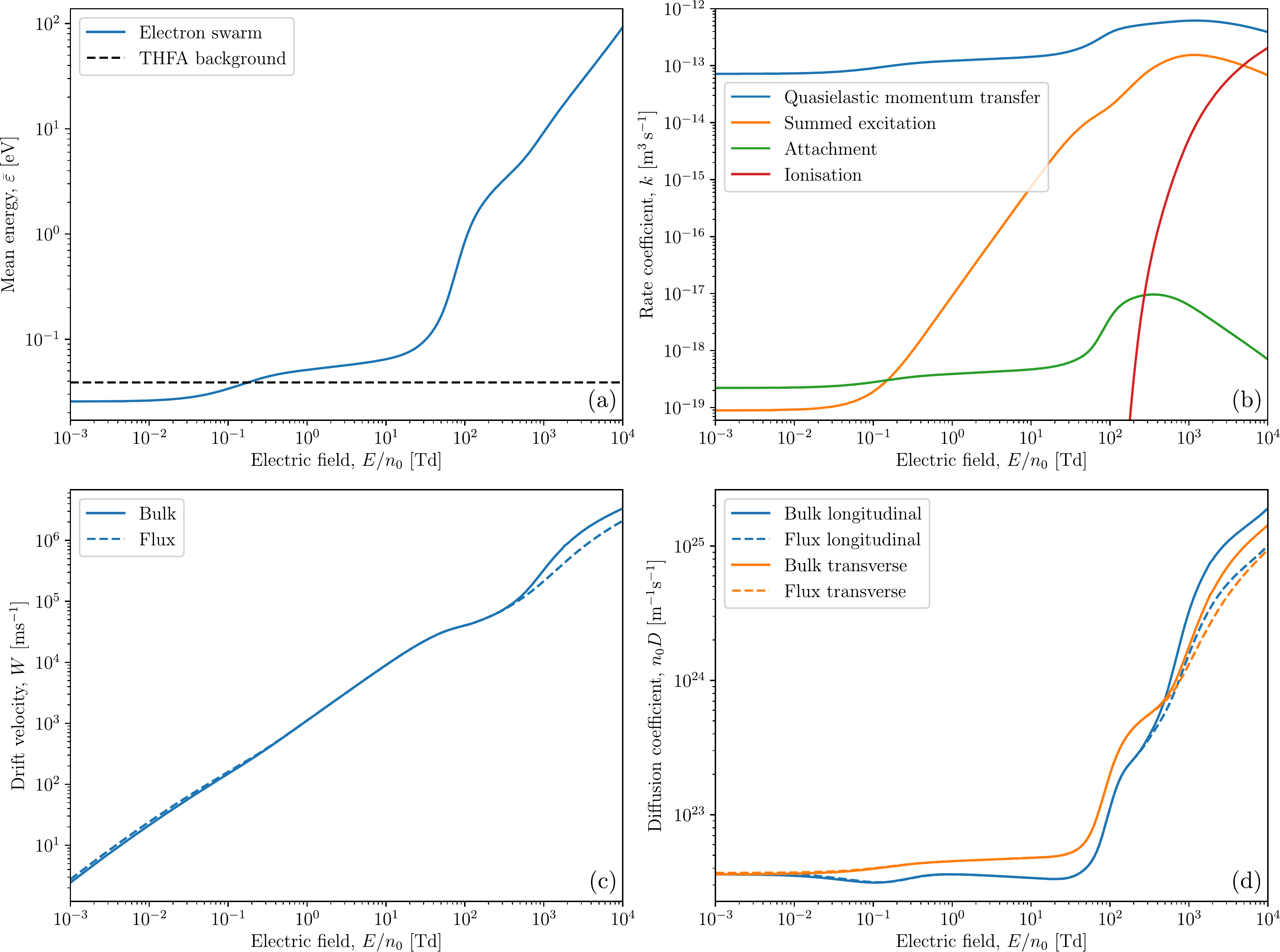}
\par\end{centering}
\caption{\label{fig:Pure}Calculated mean electron energy, (a), rate coefficients,
(b), drift velocities, (c), and diffusion coefficients, (d), for electrons
in pure THFA vapour at 300 K over a range of reduced electric fields.
See also the legends for further details.}
\end{figure}
Figure \ref{fig:Pure} shows the results of using our refined cross
section set to simulate electron swarm transport coefficients in pure
gaseous THFA at 300 K, across a range of reduced electric fields from
0.001 Td up to 10,000 Td. We employ our multi-term Boltzmann solver
here also, but find the two-term approximation to be fairly sufficient
for all but diffusion at the highest $E/n_{0}$ considered, which
can be in error by up to 59\% in the case of the bulk longitudinal
diffusion coefficient. Figure \ref{fig:Pure}(a) shows the mean energy
of the electron swarm, alongside that for the background THFA vapour
for comparison. In the low-field regime, the mean electron energy
is $\sim26\ \mathrm{meV}$, which is substantially lower than the
thermal background of $\sim39\ \mathrm{meV}$ due primarily to attachment
cooling. As $E/n_{0}$ increases, heating due to the field increases
the mean energy of the swarm to eventually reach thermal equilibrium
with the background THFA at $\sim0.175\ \mathrm{Td}$. Beyond this
$E/n_{0}$, the mean energy continues increasing monotonically, with
its ascent occasionally slowing due to the onset of the excitation
channels (around 0.5 Td) and the ionisation channel (around 200 Td).
Figure \ref{fig:Pure}(b) shows rate coefficients for quasielastic
momentum transfer, summed excitation, attachment and ionisation. The
quasielastic momentum transfer rate coefficient remains fairly constant
up until 30 Td, after which it increases slightly, likely due to the
maximum in the magnitude of the quasielastic MTCS at $\sim0.18\ \mathrm{eV}$.
The summed excitation rate coefficient starts off constant at very
low fields, but begins to increase, rather early, from $5\times10^{-2}\ \mathrm{Td}$
due to the additional opening of the vibrational excitation channels.
This increase continues monotonically until reaching a maximum at
$\sim1200\ \mathrm{Td}$, after which the summed excitation rate decreases
slightly at very high fields. The attachment rate coefficient also
starts off as being constant, before increasing monotonically to a
peak at $\sim350\ \mathrm{Td}$, likely due to the associated peaks
in the attachment cross section magnitude at $\sim2.6\ \mathrm{eV}$
and $\sim6.4\ \mathrm{eV}$. As there is no appreciable attachment
cross section beyond $\sim12\ \mathrm{eV}$, there is a corresponding
drop in the attachment rate coefficient from this point onward. The
ionisation rate coefficient is zero at low fields before increasing
monotonically and becoming appreciable from around 200 Td, onward.
Figure \ref{fig:Pure}(c) shows the bulk and flux drift velocities
of the swarm, both of which are seen to increase monotonically with
increasing $E/n_{0}$. At low fields, the flux drift velocity slightly
exceeds the bulk, indicating that electrons are being preferentially
attached at the front of the swarm and so shifting the centre of mass
towards the rear. At intermediate fields, between roughly 1 Td and
300 Td, nonconservative effects are sufficiently small such that the
bulk and flux drift velocities coincide. Above 300 Td, the bulk drift
velocity now exceeds the flux, due to ionisation preferentially creating
electrons at the front of the swarm. Figure \ref{fig:Pure}(d) shows
the bulk and flux diffusion coefficients of the swarm in both the
longitudinal and transverse directions relative to the field. As with
the drift velocities, for intermediate $E/n_{0}$ between roughly
1 Td and 300 Td, nonconservative effects are minimal and the bulk
and flux diffusion coefficients coincide. The transverse diffusion
coefficient differs the least between its flux and bulk counterparts.
At low fields, below 1 Td, there is only a very slight decrease in
the bulk transverse diffusion compared to its flux counterpart, which
we attribute to the slight preferential attachment of electrons toward
the sides of the swarm. At high fields, above 300 Td, there is a substantial
increase in bulk transverse diffusion over the flux, which we naturally
attribute to the preferential ionisation production of electrons toward
the swarm sides. The bulk longitudinal coefficient follows a similar
trend, for these same reasons of preferential attachment and ionisation
in the longitudinal direction.

\section{\label{sec:Conclusion}Conclusion}

We have formed a complete and self-consistent set of electron-THFA
cross sections, by constructing an initially-proposed set from the
literature and then refining its least-certain aspects by measuring
and analysing electron swarm transport coefficients in admixtures
of THFA in argon. Notably, this swarm analysis and cross section refinement
was performed automatically and objectively using a neural network
model, Eq. \eqref{eq:nn}, trained on cross sections from the LXCat
project \citep{Pancheshnyi2012,Pitchford2017,LXCat}. Our neural network
determined plausible cross sections for attachment and neutral dissociation,
in their entirety, from the measured swarm data, as well as cross
sections for ionisation and quasielastic momentum transfer subject
to constraints given by known experimental error bars. We subsequently
used our Boltzmann equation solver to calculate transport coefficients
for this refined cross section set, and found an improved consistency
with our experimental admixture measurements. We also calculated transport
coefficients for electrons in pure THFA, revealing the interesting
phenomenon of attachment cooling of the electron swarm below the thermal
background.

Given the similar methodology and swarm measurements between the present
investigation and our previous successful refinement of an electron-THF
cross section set \citep{Stokes2020}, we believe our refined set
of electron-THFA cross sections should be of comparable quality, if
not a little better, to one hand-fitted by an expert. As there is
evidently still some room for improvement, it is fortunate that this
machine learning approach makes it straightforward to revisit THFA
as new swarm data, cross section constraints, or LXCat training data
becomes available. 

A known limitation \citep{Stokes2019,Stokes2020} of the present machine
learning approach to swarm analysis is that it provides a unique solution
to a problem for which multiple plausible solutions are likely to
exist. In the future, we intend to address this deficiency by quantifying
the uncertainty in the predicted cross sections using a suitable alternative
machine learning model \citep{Bishop1994,Sohn2015,Mirza2014,Dinh2014,Dinh2016,Kingma2018}.
Finally, we also plan to apply our machine learning approach to determine
complete and self-consistent cross section sets for other molecules
of biological interest, including those for water \citep{White2014}. 

\ack{}{}

Thanks are due to A. Bustos and G. Bustos for their technical assistance.
The authors gratefully acknowledge the financial support of the Australian
Research Council through the Discovery Projects Scheme (Grant \#DP180101655).
JdeU thanks PAPIIT-UNAM, Project IN118520 for support. G.G. acknowledges
support from the Spanish Ministerio de Ciencia, Innovación y Universidades-MICIU
(Projects FIS2016-80440 and PID2019-104727RB-C21) and CSIC (Project
LINKA 20085).

\section*{Data Availability Statement}

The data that supports the findings of this study are available within
the article.

\appendix

\section{Pulsed-Townsend electron swarm measurement results for various admixtures
of THFA in argon}

~
\begin{table}[H]
\centering{}\caption{\label{tab:Measured-pulsed-Townsend-0.2}Measured pulsed-Townsend
electron swarm transport coefficients for a $0.2\%$ admixture of
THFA in argon. Estimated experimental uncertainties are $\pm2\%$
for drift velocities, $W$, $\pm6\%$ for effective Townsend first
ionisation coefficients, $\alpha_{\mathrm{eff}}/n_{0}$, and $\pm12\%$
for longitudinal diffusion coefficients, $n_{0}D_{L}$.}
\begin{tabular}{|c|c|c|c|}
\hline 
$E/n_{0}\ \left[\mathrm{Td}\right]$ & $W\ \left[10^{4}\ \mathrm{m}\,\mathrm{s}^{-1}\right]$ & $\alpha_{\mathrm{eff}}/n_{0}\ \left[10^{-24}\ \mathrm{m^{2}}\right]$ & $n_{0}D_{L}\ \left[10^{24}\ \mathrm{m^{-1}}\,\mathrm{s^{-1}}\right]$\tabularnewline
\hline 
\hline 
1.0 & 0.903 &  & 7.43\tabularnewline
\hline 
1.2 & 0.964 &  & 5.16\tabularnewline
\hline 
1.4 & 1.03 &  & 4.98\tabularnewline
\hline 
1.6 & 1.05 & -1.24 & 3.62\tabularnewline
\hline 
1.8 & 1.06 & -1.26 & 3.17\tabularnewline
\hline 
2.0 & 1.04 & -2.08 & 2.47\tabularnewline
\hline 
2.3 & 0.996 & -1.72 & 1.91\tabularnewline
\hline 
2.6 & 0.944 & -1.63 & 1.64\tabularnewline
\hline 
3.0 & 0.885 & -1.43 & 1.65\tabularnewline
\hline 
3.3 & 0.849 & -1.45 & 1.63\tabularnewline
\hline 
3.6 & 0.825 & -1.24 & 1.76\tabularnewline
\hline 
4.0 & 0.801 & -1.25 & 1.98\tabularnewline
\hline 
4.5 & 0.787 & -1.32 & 2.39\tabularnewline
\hline 
5.0 & 0.789 & -1.32 & 2.39\tabularnewline
\hline 
5.5 & 0.801 &  & 3.14\tabularnewline
\hline 
6.0 & 0.814 &  & 3.51\tabularnewline
\hline 
7.0 & 0.857 &  & 4.07\tabularnewline
\hline 
8.0 & 0.908 & 1.19 & 4.99\tabularnewline
\hline 
9.0 & 0.975 & 1.91 & 5.16\tabularnewline
\hline 
10 & 1.05 & 2.97 & 6.17\tabularnewline
\hline 
12 & 1.18 & 4.75 & 6.56\tabularnewline
\hline 
14 & 1.32 & 6.67 & 8.50\tabularnewline
\hline 
16 & 1.49 & 6.69 & 8.07\tabularnewline
\hline 
18 & 1.66 & 7.34 & 7.82\tabularnewline
\hline 
20 & 1.83 & 7.17 & 7.38\tabularnewline
\hline 
23 & 2.12 & 7.80 & 5.52\tabularnewline
\hline 
26 & 2.38 & 9.12 & 4.71\tabularnewline
\hline 
30 & 2.71 & 17.3 & 4.96\tabularnewline
\hline 
33 & 2.98 & 24.8 & 4.33\tabularnewline
\hline 
36 & 3.21 & 35.5 & 4.38\tabularnewline
\hline 
40 & 3.52 & 53.2 & 3.88\tabularnewline
\hline 
45 & 3.86 & 82.2 & 3.91\tabularnewline
\hline 
50 & 4.23 & 120 & 3.89\tabularnewline
\hline 
55 & 4.59 & 163 & 3.84\tabularnewline
\hline 
60 & 4.98 & 206 & 3.75\tabularnewline
\hline 
70 & 5.69 & 325 & 3.92\tabularnewline
\hline 
80 & 6.43 & 454 & 3.87\tabularnewline
\hline 
90 & 7.07 & 623 & 3.82\tabularnewline
\hline 
100 & 7.80 & 707 & 3.81\tabularnewline
\hline 
120 & 9.23 & 996 & 4.04\tabularnewline
\hline 
140 & 10.6 & 1400 & 4.45\tabularnewline
\hline 
160 & 12.0 & 1780 & 4.79\tabularnewline
\hline 
180 & 13.6 & 2070 & 4.58\tabularnewline
\hline 
200 & 14.8 & 2270 & 4.85\tabularnewline
\hline 
230 & 17.1 & 2900 & 4.57\tabularnewline
\hline 
260 & 19.7 & 3600 & 5.56\tabularnewline
\hline 
300 & 23.4 & 3890 & 5.49\tabularnewline
\hline 
330 & 25.0 & 4380 & 5.76\tabularnewline
\hline 
\end{tabular}
\end{table}
\begin{table}[H]
\centering{}\caption{Measured pulsed-Townsend electron swarm transport coefficients for
a $0.5\%$ admixture of THFA in argon. Estimated experimental uncertainties
are $\pm2\%$ for drift velocities, $W$, $\pm6\%$ for effective
Townsend first ionisation coefficients, $\alpha_{\mathrm{eff}}/n_{0}$,
and $\pm12\%$ for longitudinal diffusion coefficients, $n_{0}D_{L}$.}
\begin{tabular}{|c|c|c|c|}
\hline 
$E/n_{0}\ \left[\mathrm{Td}\right]$ & $W\ \left[10^{4}\ \mathrm{m}\,\mathrm{s}^{-1}\right]$ & $\alpha_{\mathrm{eff}}/n_{0}\ \left[10^{-24}\ \mathrm{m^{2}}\right]$ & $n_{0}D_{L}\ \left[10^{24}\ \mathrm{m^{-1}}\,\mathrm{s^{-1}}\right]$\tabularnewline
\hline 
\hline 
2.6 & 1.59 & -1.54 & 3.14\tabularnewline
\hline 
3.0 & 1.58 &  & 2.85\tabularnewline
\hline 
3.3 & 1.54 & -2.37 & 2.51\tabularnewline
\hline 
3.6 & 1.49 & -2.71 & 2.16\tabularnewline
\hline 
4.0 & 1.44 & -3.34 & 2.26\tabularnewline
\hline 
4.5 & 1.38 & -3.56 & 2.29\tabularnewline
\hline 
5.0 & 1.32 & -3.89 & 2.35\tabularnewline
\hline 
5.5 & 1.29 & -3.55 & 2.36\tabularnewline
\hline 
6.0 & 1.26 & -4.73 & 2.40\tabularnewline
\hline 
7.0 & 1.25 & -4.52 & 2.44\tabularnewline
\hline 
8.0 & 1.27 & -3.62 & 2.77\tabularnewline
\hline 
9.0 & 1.30 & -3.17 & 3.11\tabularnewline
\hline 
10 & 1.34 & -2.49 & 3.32\tabularnewline
\hline 
12 & 1.45 & -2.29 & 3.78\tabularnewline
\hline 
14 & 1.57 &  & 3.98\tabularnewline
\hline 
16 & 1.70 &  & 4.27\tabularnewline
\hline 
18 & 1.85 & 2.67 & 4.88\tabularnewline
\hline 
20 & 2.00 & 4.14 & 4.64\tabularnewline
\hline 
23 & 2.20 & 6.36 & 4.84\tabularnewline
\hline 
26 & 2.40 & 11.6 & 5.22\tabularnewline
\hline 
30 & 2.71 & 20.6 & 5.22\tabularnewline
\hline 
33 & 2.92 & 30.8 & 5.22\tabularnewline
\hline 
36 & 3.16 & 42.0 & 5.01\tabularnewline
\hline 
40 & 3.49 & 61.8 & 4.37\tabularnewline
\hline 
45 & 3.89 & 90.2 & 3.80\tabularnewline
\hline 
50 & 4.23 & 120 & 3.74\tabularnewline
\hline 
55 & 4.62 & 171 & 3.57\tabularnewline
\hline 
60 & 4.93 & 219 & 3.58\tabularnewline
\hline 
70 & 5.75 & 334 & 3.72\tabularnewline
\hline 
80 & 6.30 & 445 & 3.81\tabularnewline
\hline 
90 & 7.00 & 560 & 3.84\tabularnewline
\hline 
100 & 7.62 & 727 & 3.93\tabularnewline
\hline 
120 & 8.94 & 1020 & 4.05\tabularnewline
\hline 
140 & 10.7 & 1360 & 4.64\tabularnewline
\hline 
160 & 12.0 & 1590 & 5.10\tabularnewline
\hline 
180 & 13.2 & 2080 & 4.57\tabularnewline
\hline 
200 & 14.8 & 2300 & 4.75\tabularnewline
\hline 
230 & 17.6 & 3010 & 5.52\tabularnewline
\hline 
260 & 19.9 & 3500 & 5.06\tabularnewline
\hline 
300 & 23.4 &  & \tabularnewline
\hline 
330 & 25.4 &  & \tabularnewline
\hline 
\end{tabular}
\end{table}
\begin{table}[H]
\centering{}\caption{Measured pulsed-Townsend electron swarm transport coefficients for
a $1\%$ admixture of THFA in argon. Estimated experimental uncertainties
are $\pm2\%$ for drift velocities, $W$, $\pm6\%$ for effective
Townsend first ionisation coefficients, $\alpha_{\mathrm{eff}}/n_{0}$,
and $\pm12\%$ for longitudinal diffusion coefficients, $n_{0}D_{L}$.}
\begin{tabular}{|c|c|c|c|}
\hline 
$E/n_{0}\ \left[\mathrm{Td}\right]$ & $W\ \left[10^{4}\ \mathrm{m}\,\mathrm{s}^{-1}\right]$ & $\alpha_{\mathrm{eff}}/n_{0}\ \left[10^{-24}\ \mathrm{m^{2}}\right]$ & $n_{0}D_{L}\ \left[10^{24}\ \mathrm{m^{-1}}\,\mathrm{s^{-1}}\right]$\tabularnewline
\hline 
\hline 
5.5 & 1.92 & -3.65 & 2.56\tabularnewline
\hline 
6.0 & 1.88 & -3.32 & 2.70\tabularnewline
\hline 
7.0 & 1.80 & -4.28 & 2.54\tabularnewline
\hline 
8.0 & 1.78 & -3.66 & 2.70\tabularnewline
\hline 
9.0 & 1.76 & -3.78 & 2.75\tabularnewline
\hline 
10 & 1.76 & -2.78 & 2.75\tabularnewline
\hline 
12 & 1.80 & -2.76 & 3.04\tabularnewline
\hline 
14 & 1.89 & -2.39 & 3.38\tabularnewline
\hline 
16 & 1.98 &  & \tabularnewline
\hline 
18 & 2.10 &  & \tabularnewline
\hline 
20 & 2.18 & 3.03 & 4.23\tabularnewline
\hline 
23 & 2.38 & 4.15 & 4.76\tabularnewline
\hline 
26 & 2.56 & 8.58 & 4.76\tabularnewline
\hline 
30 & 2.83 & 16.0 & 4.73\tabularnewline
\hline 
33 & 3.05 & 27.3 & 4.56\tabularnewline
\hline 
36 & 3.28 & 36.5 & 4.24\tabularnewline
\hline 
40 & 3.54 & 55.0 & 4.28\tabularnewline
\hline 
45 & 3.96 & 80.9 & 3.94\tabularnewline
\hline 
50 & 4.25 & 113 & 3.83\tabularnewline
\hline 
55 & 4.62 & 161 & 3.74\tabularnewline
\hline 
60 & 5.04 & 215 & 3.62\tabularnewline
\hline 
70 & 5.70 & 320 & 3.63\tabularnewline
\hline 
80 & 6.31 & 454 & 3.75\tabularnewline
\hline 
90 & 7.00 & 603 & 3.90\tabularnewline
\hline 
100 & 7.65 & 767 & 4.24\tabularnewline
\hline 
120 & 9.05 & 1060 & 4.64\tabularnewline
\hline 
140 & 10.4 & 1360 & 4.43\tabularnewline
\hline 
160 & 11.8 & 1550 & 4.46\tabularnewline
\hline 
180 & 13.4 & 2010 & 4.44\tabularnewline
\hline 
200 & 14.8 & 2390 & 4.54\tabularnewline
\hline 
230 & 16.9 & 3110 & 4.77\tabularnewline
\hline 
260 & 19.9 & 3560 & 5.46\tabularnewline
\hline 
300 & 22.9 &  & \tabularnewline
\hline 
330 & 25.6 &  & \tabularnewline
\hline 
360 & 28.2 &  & \tabularnewline
\hline 
400 & 30.4 &  & \tabularnewline
\hline 
\end{tabular}
\end{table}
\begin{table}[H]
\centering{}\caption{Measured pulsed-Townsend electron swarm transport coefficients for
a $2\%$ admixture of THFA in argon. Estimated experimental uncertainties
are $\pm2\%$ for drift velocities, $W$, $\pm6\%$ for effective
Townsend first ionisation coefficients, $\alpha_{\mathrm{eff}}/n_{0}$,
and $\pm12\%$ for longitudinal diffusion coefficients, $n_{0}D_{L}$.}
\begin{tabular}{|c|c|c|c|}
\hline 
$E/n_{0}\ \left[\mathrm{Td}\right]$ & $W\ \left[10^{4}\ \mathrm{m}\,\mathrm{s}^{-1}\right]$ & $\alpha_{\mathrm{eff}}/n_{0}\ \left[10^{-24}\ \mathrm{m^{2}}\right]$ & $n_{0}D_{L}\ \left[10^{24}\ \mathrm{m^{-1}}\,\mathrm{s^{-1}}\right]$\tabularnewline
\hline 
\hline 
10 & 2.55 & -13.0 & 2.45\tabularnewline
\hline 
12 & 2.48 & -12.8 & 2.57\tabularnewline
\hline 
14 & 2.49 & -12.7 & 2.73\tabularnewline
\hline 
16 & 2.53 & -11.3 & 2.93\tabularnewline
\hline 
18 & 2.59 & -9.13 & 3.03\tabularnewline
\hline 
20 & 2.68 & -8.29 & 3.06\tabularnewline
\hline 
23 & 2.83 & -4.70 & 3.20\tabularnewline
\hline 
26 & 2.99 & -1.81 & 3.35\tabularnewline
\hline 
30 & 3.23 & 7.66 & 3.46\tabularnewline
\hline 
33 & 3.42 & 14.6 & 3.48\tabularnewline
\hline 
36 & 3.58 & 22.1 & 3.49\tabularnewline
\hline 
40 & 3.84 & 36.0 & 3.55\tabularnewline
\hline 
45 & 4.16 & 59.6 & 3.64\tabularnewline
\hline 
50 & 4.48 & 92.1 & 3.71\tabularnewline
\hline 
55 & 4.79 & 132 & 3.79\tabularnewline
\hline 
60 & 5.15 & 177 & 3.98\tabularnewline
\hline 
70 & 5.79 & 297 & 3.92\tabularnewline
\hline 
80 & 6.52 & 449 & 3.94\tabularnewline
\hline 
90 & 7.01 & 556 & 3.89\tabularnewline
\hline 
100 & 7.90 & 661 & 4.10\tabularnewline
\hline 
120 & 9.08 & 950 & 4.44\tabularnewline
\hline 
140 & 10.6 & 1200 & 4.40\tabularnewline
\hline 
160 & 12.1 & 1660 & 4.72\tabularnewline
\hline 
180 & 13.4 & 1930 & 5.08\tabularnewline
\hline 
200 & 15.0 & 2470 & 4.79\tabularnewline
\hline 
230 & 17.8 & 2860 & 5.63\tabularnewline
\hline 
260 & 19.4 & 3340 & 5.06\tabularnewline
\hline 
300 & 23.6 &  & \tabularnewline
\hline 
330 & 26.0 &  & \tabularnewline
\hline 
360 & 28.7 &  & \tabularnewline
\hline 
400 & 32.1 &  & \tabularnewline
\hline 
450 & 34.5 &  & \tabularnewline
\hline 
\end{tabular}
\end{table}
\begin{table}[H]
\centering{}\caption{\label{tab:Measured-pulsed-Townsend-5}Measured pulsed-Townsend electron
swarm transport coefficients for a $5\%$ admixture of THFA in argon.
Estimated experimental uncertainties are $\pm2\%$ for drift velocities,
$W$, $\pm6\%$ for effective Townsend first ionisation coefficients,
$\alpha_{\mathrm{eff}}/n_{0}$, and $\pm12\%$ for longitudinal diffusion
coefficients, $n_{0}D_{L}$.}
\begin{tabular}{|c|c|c|c|}
\hline 
$E/n_{0}\ \left[\mathrm{Td}\right]$ & $W\ \left[10^{4}\ \mathrm{m}\,\mathrm{s}^{-1}\right]$ & $\alpha_{\mathrm{eff}}/n_{0}\ \left[10^{-24}\ \mathrm{m^{2}}\right]$ & $n_{0}D_{L}\ \left[10^{24}\ \mathrm{m^{-1}}\,\mathrm{s^{-1}}\right]$\tabularnewline
\hline 
\hline 
26 & 3.89 & -21.4 & 2.88\tabularnewline
\hline 
30 & 4.03 & -9.54 & 2.99\tabularnewline
\hline 
33 & 4.16 & -9.47 & 3.01\tabularnewline
\hline 
36 & 4.29 & -2.91 & 3.09\tabularnewline
\hline 
40 & 4.55 & 3.40 & 3.21\tabularnewline
\hline 
45 & 4.84 & 15.7 & 3.36\tabularnewline
\hline 
50 & 5.11 & 30.6 & 3.41\tabularnewline
\hline 
55 & 5.40 & 63.0 & 3.40\tabularnewline
\hline 
60 & 5.73 & 99.1 & 3.37\tabularnewline
\hline 
70 & 6.30 & 178 & 3.48\tabularnewline
\hline 
80 & 6.88 & 306 & 3.69\tabularnewline
\hline 
90 & 7.48 & 438 & 3.97\tabularnewline
\hline 
100 & 8.10 & 582 & 4.10\tabularnewline
\hline 
120 & 9.49 & 918 & 4.25\tabularnewline
\hline 
140 & 10.9 & 1260 & 4.82\tabularnewline
\hline 
160 & 12.4 & 1630 & 5.07\tabularnewline
\hline 
180 & 13.5 & 2100 & 5.14\tabularnewline
\hline 
200 & 15.3 & 2350 & 4.75\tabularnewline
\hline 
230 & 18.1 & 2770 & 5.57\tabularnewline
\hline 
260 & 19.8 & 3330 & 5.07\tabularnewline
\hline 
330 & 26.5 &  & \tabularnewline
\hline 
\end{tabular}
\end{table}

\bibliographystyle{iopart-num}
\addcontentsline{toc}{section}{\refname}\bibliography{references}

\end{document}